%% file: paper.tex
\pdfoutput=1
\documentclass[10pt, conference, compsocconf]{IEEEtran}

\usepackage{booktabs} 

\usepackage{graphicx}
\usepackage{grffile} 
\usepackage{subfigure}
\usepackage{xspace}
\usepackage[ruled, vlined, linesnumbered]{algorithm2e}
\usepackage{comment}
\usepackage{url}
\usepackage{multirow}
\usepackage{hyperref}
\usepackage{xcolor}
\usepackage{tabularx}
\usepackage{pbox}
\usepackage{enumitem,kantlipsum}

\newboolean{draft}
\setboolean{draft}{true}

\ifthenelse{\boolean{draft}}{

}{}

\newcommand{\myfigure}[1]{Figure~#1}
\newcommand{\mytable}[1]{Table~#1}

\newcommand{\ie}{\textit{i}.\textit{e}., }
\newcommand{\eg}{\textit{e}.\textit{g}., }

\newcommand{\quotes}[1]{``#1''}
\newcommand{\nbo}[1]{\emph{N\"aive BO}}
\newcommand{\abo}[1]{\emph{Augmented BO}}

\renewcommand\figurename{Figure} 

\begin{document}
\title{Low-Level Augmented Bayesian Optimization for Finding the Best Cloud VM}

\author{
\IEEEauthorblockN{
Chin-Jung Hsu,
Vivek Nair,
Vincent W. Freeh,
Tim Menzies}
\IEEEauthorblockA{
Department of Computer Science, North Carolina State University\\
chsu6@ncsu.edu, vivekaxl@gmail.com, vwfreeh@ncsu.edu, tim.menzies@gmail.com}
}

\maketitle

\input{abstract}


\IEEEpeerreviewmaketitle

\input{introduction.tex}
\input{background.tex}

\input{bo.tex}
\input{approach.tex}

\input{evaluation.tex}
\input{relatedwork.tex}
\input{conclusion.tex}


\bibliographystyle{plain}
\input{reference.bbl}               


\end{document}

%% file: abstract.tex
\begin{abstract}

With the advent of big data applications, which tends to have longer execution time,
choosing the right cloud VM to run these applications
has significant performance as well as economic implications.
For example, in our large-scale empirical study of 107 different workloads
on three popular big data systems,
we found that a wrong choice
can lead to a 20 times slowdown or an increase in cost by 10 times.

Bayesian optimization is a technique for optimizing
expensive (black-box) functions.
Previous attempts have only used
instance-level information (such as \# of cores, memory size)
which is not sufficient to represent the search space.
In this work, we discover that this may lead to
the \emph{fragility} problem---either
incurs high search cost or finds only the sub-optimal solution.
The central insight of this paper is to use low-level performance information
to augment the process of Bayesian Optimization.
Our novel low-level augmented Bayesian Optimization is rarely worse than
current practices and often performs much better (in 46 of 107 cases).
Further, it significantly reduces the search cost
in nearly half of our case studies.

Based on this work, we conclude that it is often insufficient to use
general-purpose off-the-shelf methods for configuring cloud instances
without augmenting those methods with essential systems knowledge such as
CPU utilization, working memory size and I/O wait time.

\end{abstract}

%% file: introduction.tex
\section{Introduction}
\label{sec:introduction}

\noindent\textbf{Motivation.} Cloud computing is a cost-effective
alternative to on-premise computing.
To accommodate diverse workloads, cloud service providers
(Amazon, Google, and Azure) offer over 100 virtual machines (VM) types~\cite{Yadwadkar2017}.
Our experiments show the optimal choice of VMs can be up to
20 times faster and 10 times less expensive than the worst VM for the same workload.
Therefore,
choosing the right VM type for a workload is essential to provide quality service while being commercially competitive~\cite{Frey2013, Yao2017}.


In this paper, we address the problem of finding a suitable cloud
VM type for a recurring job.
This problem is further aggravated by
the \textit{long execution times of the workloads} since a brute-force approach will no longer be a viable option.
Furthermore, because there are charges to evaluate, this
\emph{decision space} must be explored efficiently.
The prior work in this area, solved this problem using two different approaches namely (1)
\emph{PARIS}~\cite{Yadwadkar2017} builds a complex performance model (using
large-scale one-time benchmark data) to predict workload performance, and
(2) \emph{CherryPick}~\cite{Alipourfard2017} uses Bayesian optimization to find the best cloud configuration.
We prefer the Bayesian Optimization (BO) method because
it does not require additional historical training data and
supports any objective functions (essential for diverse workloads).


\begin{figure}
    \centering
    \includegraphics[width=0.45\textwidth]{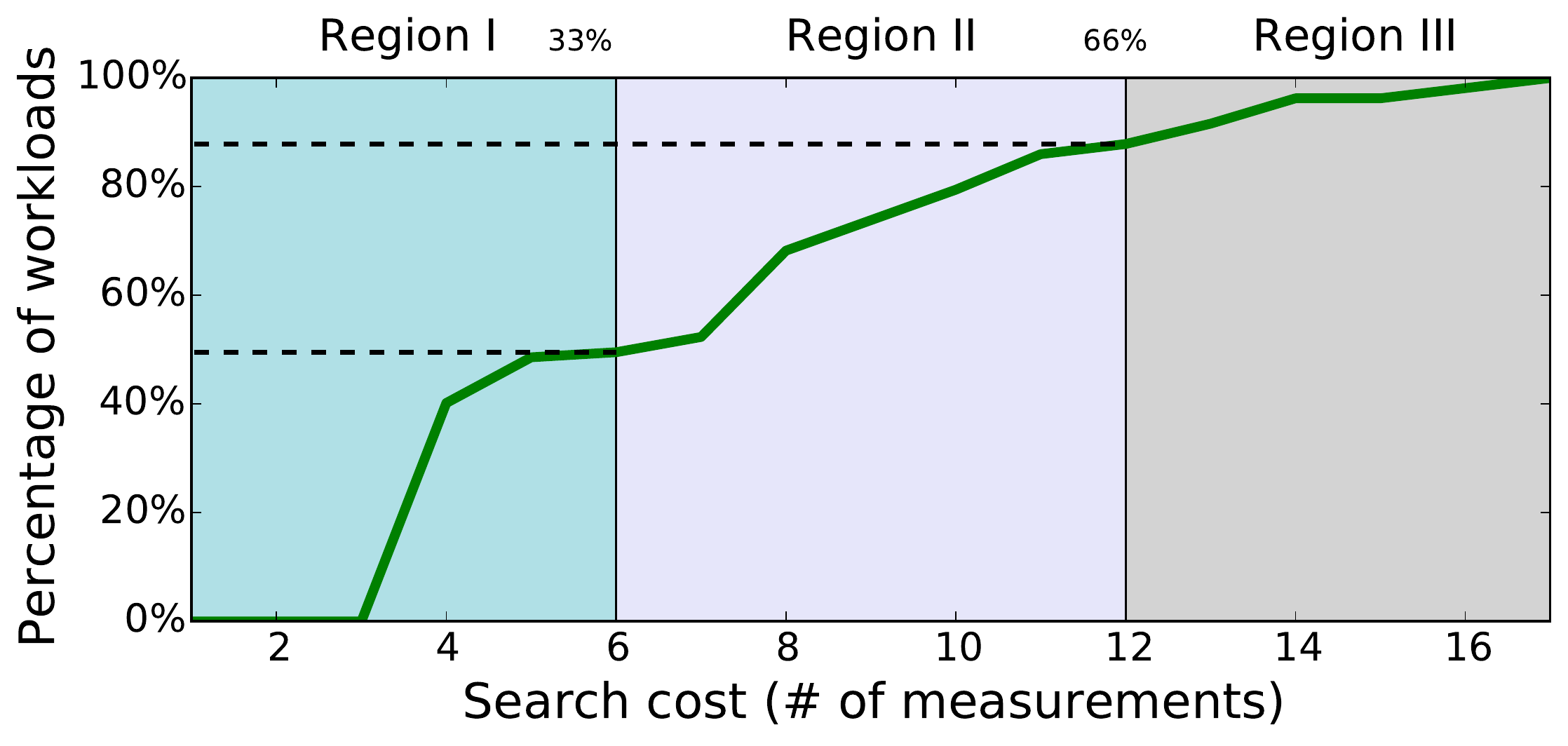}
    \vspace*{-2mm}
    \caption{The number of measurements required by Bayesian Optimization (as used in ~\cite{Alipourfard2017}) to find the optimal VM type. We observe that 50\% and 85\% of the workloads (shown in dashed lines) require 6 (33\% of the search space) and 12 (66\% of the search space) measurements respectively. Bayesian Optimization is not always effective for any workload.
    The fragility problem---either incurs high search cost or yields sub-optimal solution (as in \emph{Region II} and \emph{Region III}.}
    \label{fig:cherrypick_issue1}
    \vspace*{-6mm}
\end{figure}

\begin{figure}
    \centering
    \includegraphics[width=0.45\textwidth]{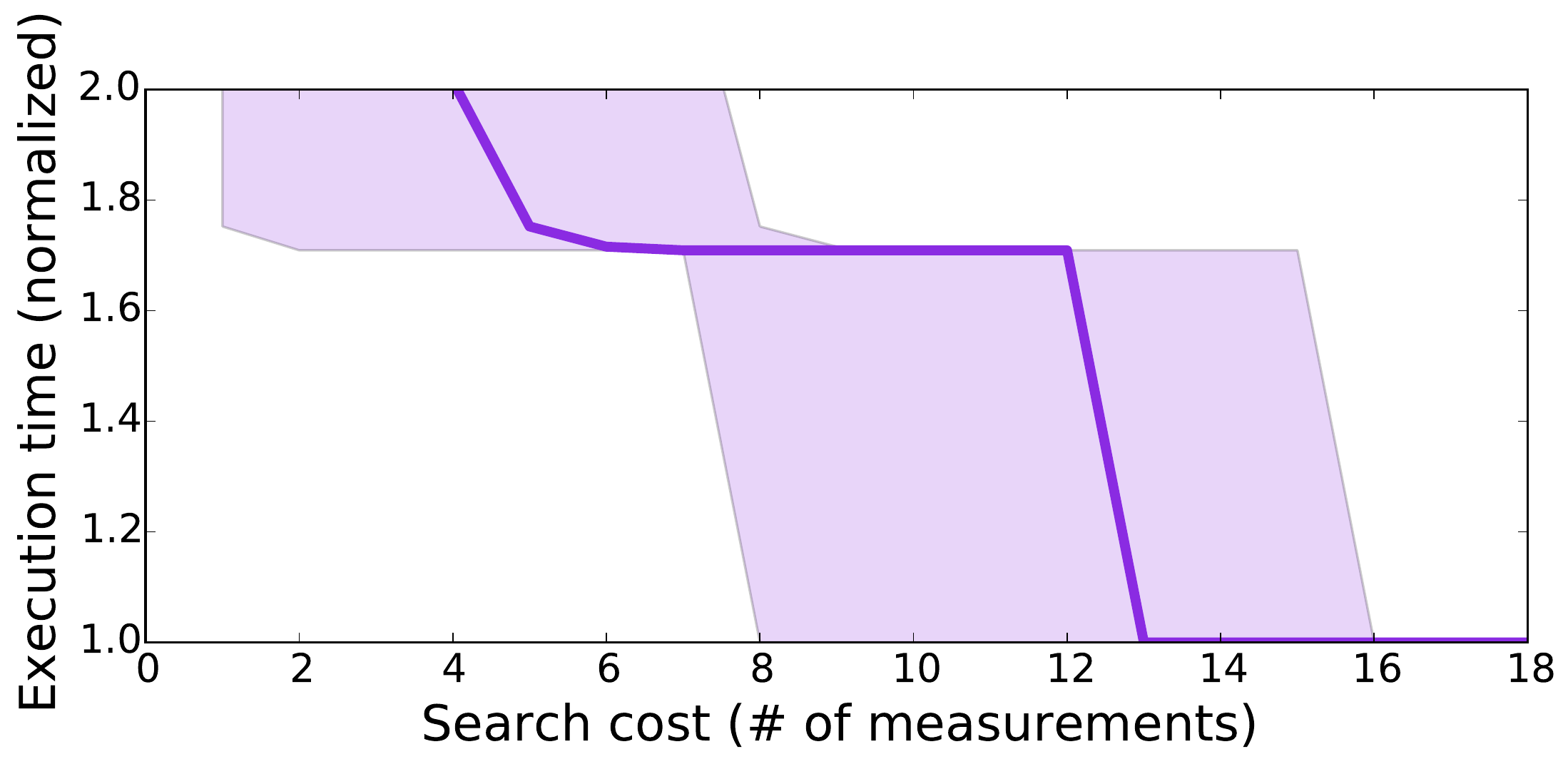}
    \vspace*{-2mm}
    \caption{Using Bayesian Optimization to find the best VM type for running the ALS algorithm on Spark.
    The horizontal axis represents the search cost, and the vertical axis represents the execution time of the workload (both are lower the better).
    A naive Bayesian Optimization method progresses slowly towards the optimal VM type.
    The low-level augmented BO method alleviates the fragility problem as shown in Figure~\ref{fig:convergence_time_1}.
    }
    \label{fig:cherrypick_issue2}
    \vspace*{-6mm}
\end{figure}

However, we have come across workloads where a BO method is ineffective---surprisingly,
we found this problem in a large number of workloads.
Our large-scale empirical study, as shown in \myfigure{~\ref{fig:cherrypick_issue1}},
reveals that BO incurs different search cost on different workloads.
We observe that BO is effective in 50\% of the workloads (in \emph{Region I}) since it requires exploration of only
33\% of the total search space.
However, we also notice that BO is not as effective at finding the optimal VM type
for the other workloads (in \emph{Region II} and \emph{Region III}).
This poor performance can be attributed to
the insufficient information (for example \# of cores, memory, etc.) used by BO during the search process.
Such VM characteristics are not sufficient to capture application behavior~\cite{Hsu2016, Yadwadkar2017, Dalibard2017}.
Consequently, BO may fail to find the optimal VM for some workloads efficiently.
Figure~\ref{fig:cherrypick_issue2} shows
how BO is sluggish to find a `better' VM type for a workload from the \emph{Region III}.
In summary, the lesson that we learned from the large-scale empirical study is \textit{BO is not a silver bullet to find optimal VM type} for any workloads.
Furthermore, it can be \textit{fragile}---either incurs higher search cost or yields a sub-optimal solution.
Without further investigation, it is hard to claim BO is an effective method for finding the best VM type.

\noindent\textbf{Our Work.}
To further understand the fragility of Bayesian Optimization,
we conducted a large-scale empirical study with three popular big data systems along with 107 different workloads and 18 different VM types (for more details refer to Section~\ref{sec:datacollection}).  
We first observe that using rule-of-thumbs (intuitions) to select the best VM type is far from ideal.
There does not exist one such best VM type for all the workloads.
Second, the same application with different input sizes may favor different VM types.
Last,
while the execution time tends to decrease with a more powerful
VM, the cost per unit time goes up, which compresses the
deployment costs.
This creates a \textit{level playing field}---several inferior
configurations in execution time are now competitive in deployment cost.
These reasons make the problem of selecting the best VM for any given workload challenging.

To find the best VM type, \emph{CherryPick}~\cite{Alipourfard2017} uses Bayesian Optimization, which sequentially evaluates the VMs and
moves closer to the optimal VM type.
As presented before, a BO method can encounter the fragility problem.
As shown in Figure~\ref{fig:cherrypick_issue2}, 
the performance of instance found after the fifth iteration is 1.75 times slower when
compared to the optimal instance type.
In this case, BO did not find the optimal solution until the thirteenth attempt.
We argue that the fragility of BO arises from the \textit{insufficient information}.
That is, characteristics of a VM such as CPU speed, core counts, memory per core and disk capacity, is not sufficient to predict its performance.
Besides, the \textit{choice of the kernel function} (the prior) and
the selection of the initial measurements are both critical to the effectiveness of BO~\cite{Brochu2010, Snoek2012, Dewancker2015, Shahriari2016}.
We believe they are also related to the fragility problem.

Low-level performance metrics are a good proxy for estimating
application and system performance~\cite{Hsu2016, Yadwadkar2017}.
They are also useful to identify performance anomalies
~\cite{Bodik2010, Novakovic2013}.
We argue that \textit{low-level performance
  information} such as I/O wait and memory usage better captures
application behavior and better
guides a BO method through the search process.


In this paper, we proposed
a novel method to augment Bayesian Optimization
by leveraging low-level performance information.
However, embedding low-level performance information is tricky since the (low-level) information is not available until the workload is executed on a given VM type.
Our proposed modeling technique seamlessly integrates
the high-level features with the low-level performance information.
The prediction model estimates the workload performance
in VMs (not measured) using the
low-level performance information collected from previous measurements.
Throughout the search process, the model keeps updating the belief
based on the new measurements.

The proposed low-level augmented Bayesian Optimization (Augmented BO) outperforms
the naive Bayesian Optimization (Naive BO)~\cite{Alipourfard2017}.
Our evaluation shows a reduction in search cost on 46 out of 107 applications
in search for the most cost-effective configuration.
Our method reduces about 20\% search cost on average for cases
with the fragility issue, and reaches 43\% reduction for some
while maintaining the same or slightly better performance
in comparison to Naive BO.

\noindent\textbf{Summary and contributions.} Our key contributions are:
\begin{enumerate}[leftmargin=*]
\item A large-scale empirical study to analyze the performance of Bayesian Optimization
on a wide range of realistic data analytics and machine learning workloads
(Section~\ref{sec:background});
\item A demonstration of fragility of BO to find the suitable instance for a specific workload (Section~\ref{sec:bo});
\item A novel low-level augmented Bayesian Optimization method to alleviate the fragility problem (Section~\ref{sec:approach}).

\end{enumerate}

%% file: background.tex
\section{Background and Motivation}
\label{sec:background}

In this section, we present the challenges of selecting the best VM type.
We also formulate our problem setting and explain why
search-based optimization is more desirable.

\subsection{Problem Formalization}

A cloud service provider presents its user with several choices of VM types ($\mathit{VM}$). Let $\mathit{VM}_i$ indicate the $i^{th}$ VM type in the list of VMs, which takes value from a finite domain $Dom(\mathit{VM}_i)$. In general, $\mathit{VM}_i$ indicates the published characteristics of VMs (such as memory size, \# of cores).
$\mathit{VM}_{ij}$ represents the $j^{th}$ characteristic of the $i^{th}$ VM type. The \textit{instance space} is thus $Dom(\mathit{VM}_1) \times Dom(\mathit{VM}_2) \times ... \times Dom(\mathit{VM}_n)$, which is the Cartesian product of the domains, where $n = \left\vert{\mathit{VM}}\right\vert$ is the number of VMs provided by the cloud service provider. 
When a workload ($w\in W$) is run on a VM ($\mathit{VM}_i$), the low-level metrics ($l_{i,w}\in L$) can be collected from the VM.
Each VM type ($\mathit{VM}$) has a corresponding performance measure $y\in Y$ (\eg{time or cost}).
We denote the performance measure associated with a given VM type and a workload by $y_{i,w}=f(\mathit{VM}_{i,w})$. In this setting, $\mathit{VM}_{i,w}$ and $y_{i,w}$  is called independent and dependent variable respectively.

\noindent Our goal is design a search method to:
\begin{enumerate}[leftmargin=*]
    \item \textit{Minimize} performance difference between
the \emph{best} VM  ($\mathit{VM^*}$) (found by search) and the optimal VM  ($\mathit{VM}^{opt})$. We find $\mathit{VM^*}$ both in terms of \textit{execution time} and \textit{deployment cost};
    \item \textit{Minimize} search cost---the number of measurements required to find the (near) optimal configuration.
\end{enumerate}
\input{tables/dataset}

\subsection{Large-scale evaluation on AWS}
\label{sec:datacollection}

To evaluate workload performance on different VMs,
we conducted a large-scale evaluation using different workloads and software systems on Amazon Web Services (AWS)~\cite{aws}.
We choose Apache Hadoop (version 2.7) and Apache Spark (version 1.5 and 2.1) as our software system~\cite{hadoop, spark}.
Our evaluation includes data processing, OLAP queries, and machine learning, which are popular workloads on Hadoop and Spark.
We choose 18 VMs and run the 30 workloads on them. \mytable{\ref{tab:dataset}} lists all the software systems and
workloads. See Section \ref{sec:evaluation} for details.

We also vary the input size or input parameters to the workloads because workload behavior may change dramatically~\cite{Dalibard2017}.
Consequently, the optimal VM type for a given workload with different inputs might also change.
By running workloads with different data sizes, we can observe whether a particular VM can sustain increasing resource requirements (of a workload).
Our motivation (for the large-scale study) was to diversify the workloads such that we can extensively benchmark VMs.
In this study, each workload is tested with three different inputs sizes.
Some tests failed because smaller VM instances run out of memory.
Those are excluded in our data set.
In total, we measure the performance and collect the low-level information of 107 workloads on 18 different VM types.

\subsection{Choosing the best VM is troublesome}
\label{sec:challenges}

Finding the best VM is often very challenging.
The growing complexity comes from five factors.

\noindent {\textbf{The increasing number of VM types:}}
To accommodate the growing number of workloads,
cloud service providers frequently adds new VM types to their already large VM portfolio. AWS, for instance, has a significant upgrade on its service two times a month on average~\cite{ec2history}.
As of December 2017, AWS provides 71 active VM types.
Such a trend would make a brute-force search for the best VM type expensive. Also, it is difficult to model the performance of a workload for distinct VM types~\cite{Yadwadkar2017}.

\noindent {\textbf{Official recommendation is insufficient:}}
AWS recommends VM types for workloads.
Even though such recommendations are beneficial for the users, these recommendations cannot be trusted completely. 
For example, users are encouraged to choose
compute-optimized VMs for CPU-intensive workloads and
memory-optimized VMs for workloads requiring large memory.
However, characterizing workloads is still considered difficult and requires expertise, which is often very expensive and sometimes unavailable. This problem is exacerbated by workloads, which regularly exercise resource components in a non-uniform manner~\cite{Ousterhout2017}.
Furthermore, it is difficult to understand the resource requirement of
a workload for achieving a specific performance objective~\cite{Yadwadkar2017}.

\begin{figure}[t]
 \centering
 \subfigure[Execution time]{
 \label{fig:motivation_bad_choice_time}
 \includegraphics[width=.22\textwidth]{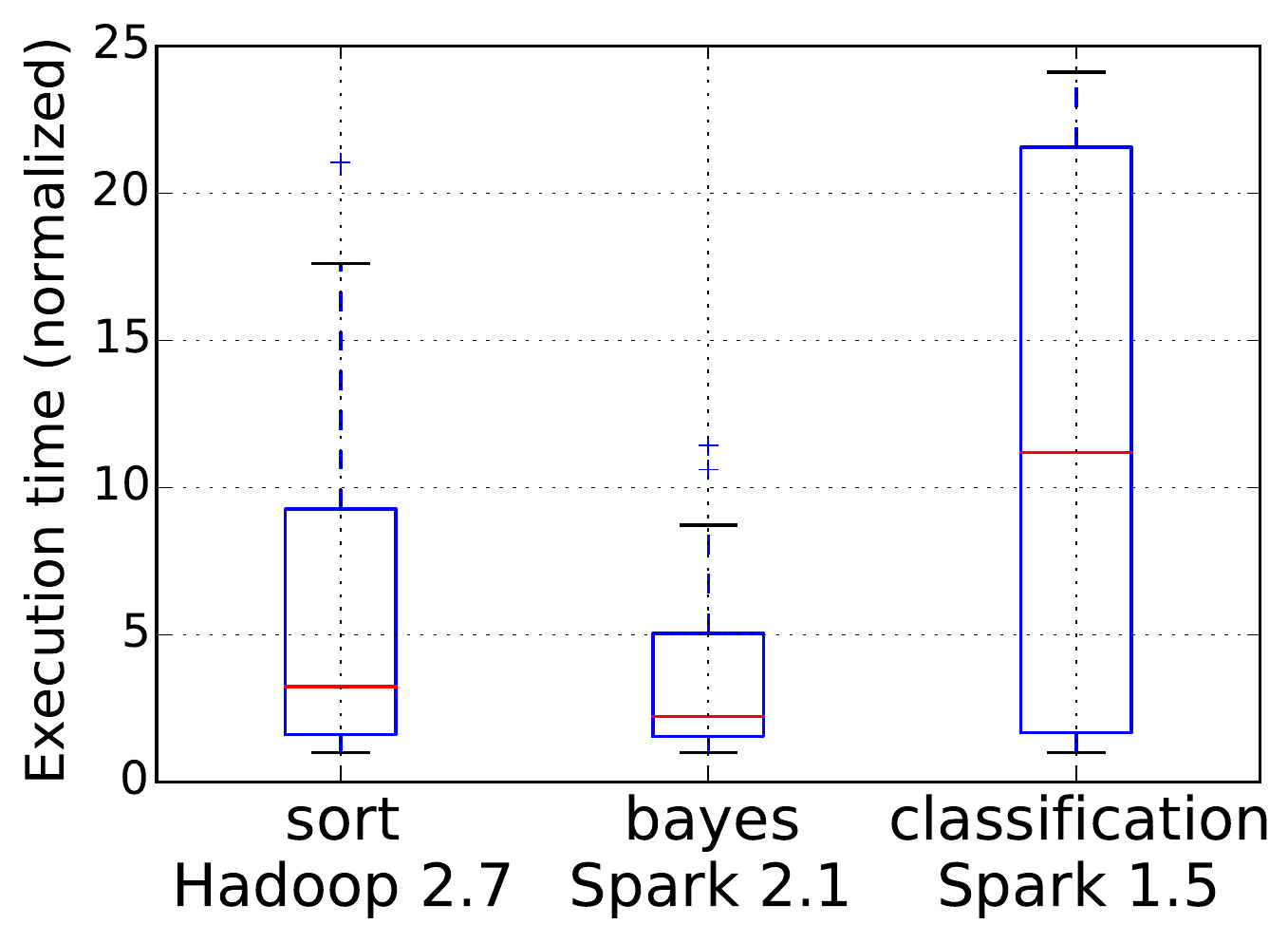}
 }
 \subfigure[Deployment cost]{
 \label{fig:motivation_bad_choice_cost}
 \includegraphics[width=.22\textwidth]{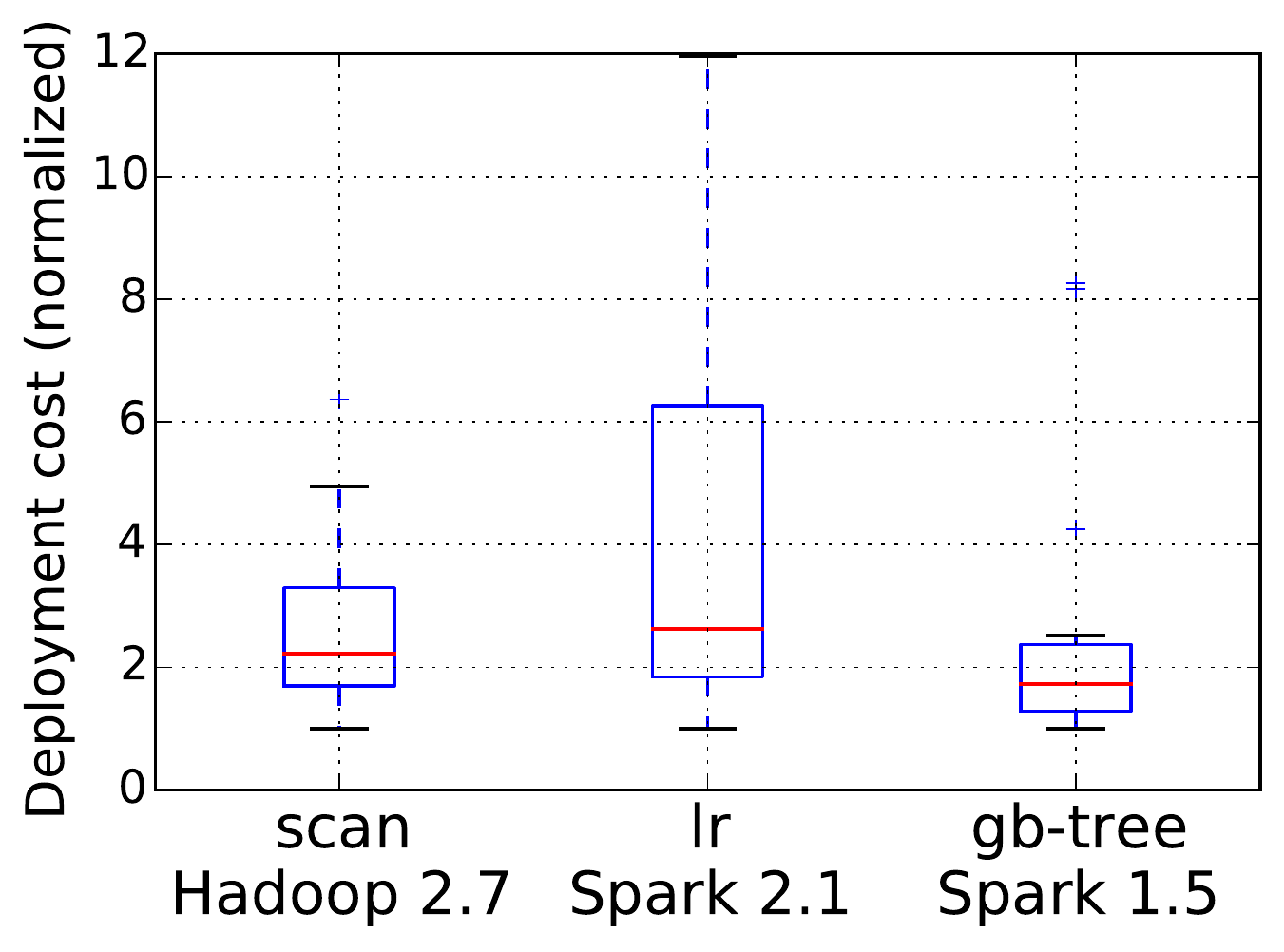}
 }
 \centering
 \caption{The execution time and deployment cost of workloads running on 18 virtual machines (different types). The execution time of  classification-Spark 1.5 using worst VM type is 20 times slower that the best VM type. Similarly the deployment cost of running Linear Regression the worst VM type is 10 times more expensive than the best VM type.}
 \label{fig:motivation_bad_choice}
 \vspace*{-4mm}
\end{figure}

\noindent {\textbf{No VM rules all:}}
Our empirical data, as shown in \myfigure{\ref{fig:motivation_bad_choice}}, demonstrates that
a bad choice can increase the execution time (of a workload) up to 20 times and can be ten times more costly than the optimal one.
Prior work reports similar results~\cite{Alipourfard2017, Yadwadkar2017}. Careless selection can often end up
with high deployment cost and longer (sub-optimal) execution time.

\begin{figure}[t]
 \centering
 \subfigure[Execution time on the most expensive VM types.]{
 \label{fig:motivation_variance_time}
 \includegraphics[width=.22\textwidth]{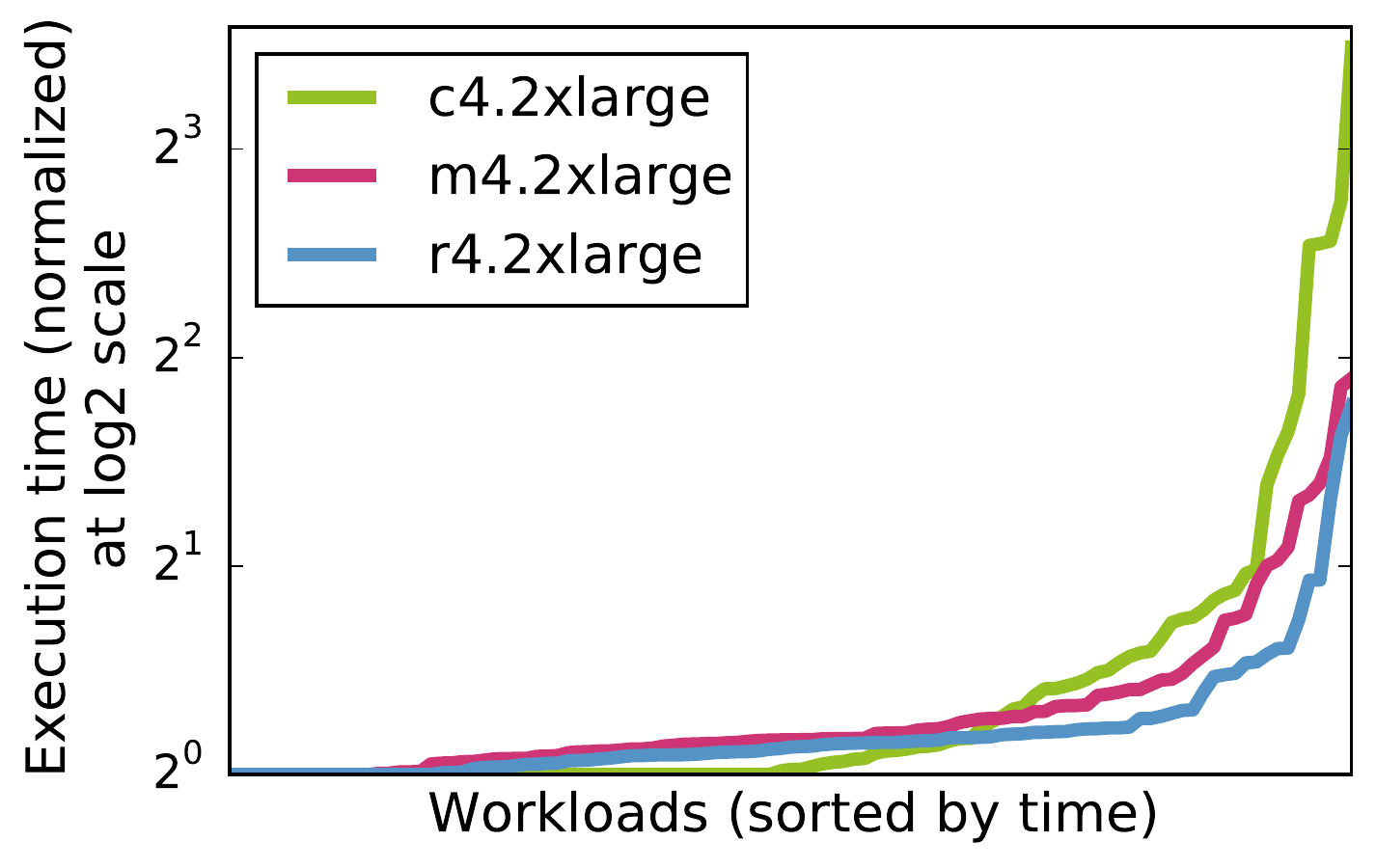}
 }
 \subfigure[Deployment cost on the least expensive VM types.]{
 \label{fig:motivation_variance_cost}
 \includegraphics[width=.22\textwidth]{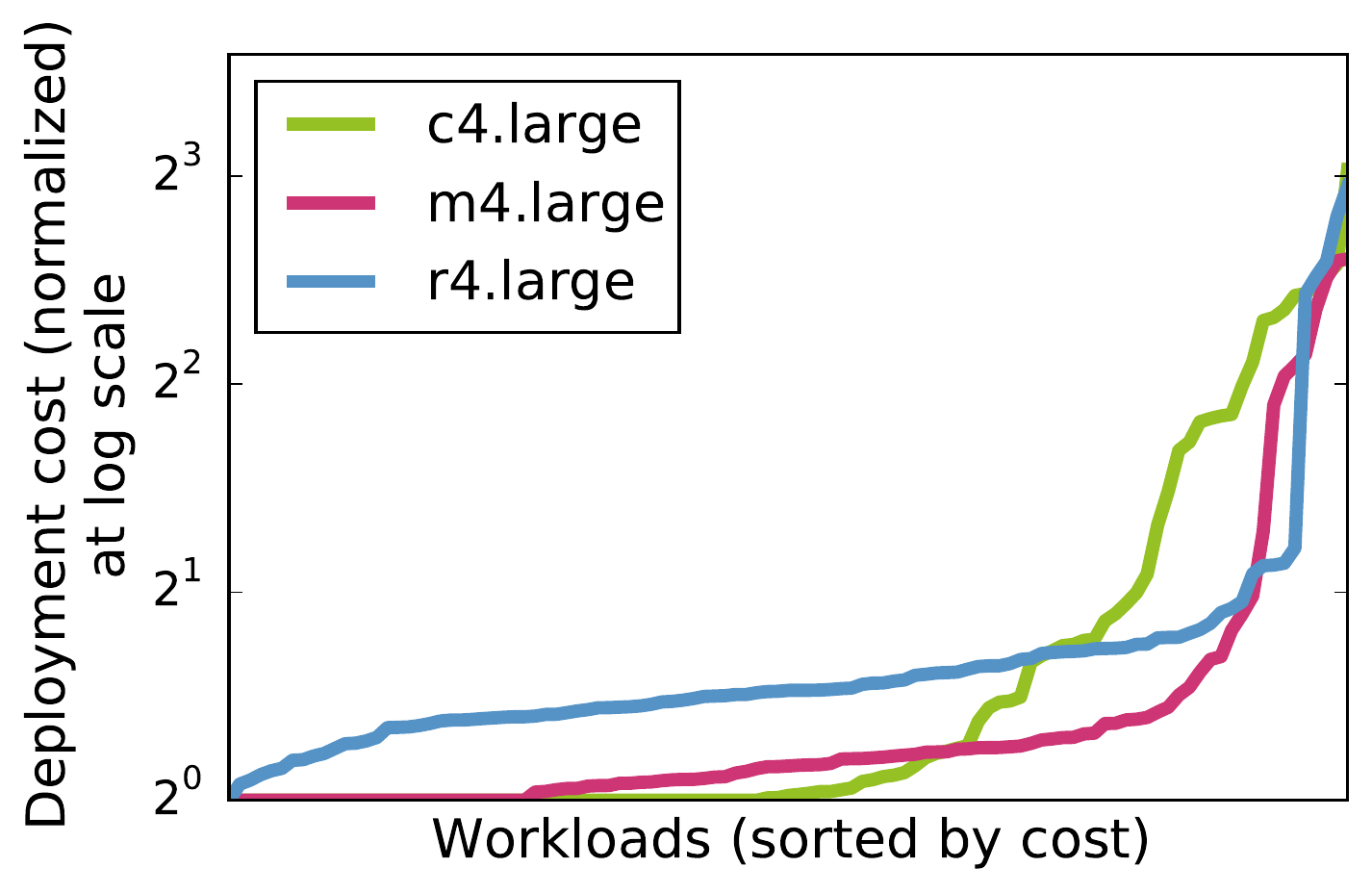}
 }
 \centering
 \caption{Performance distribution over different workloads. The performance is normalized to the optimal performance measured in the 18 virtual machine. The \emph{x-axis} represents workloads, sorted by their normalized performance.  Both choosing the most expensive and the cheapest VM types are not desirable.}
 \label{fig:motivation_variance}
 \vspace*{-6mm}
\end{figure}

Even though users are willing to pay a higher cost in exchange for performance, choosing the most expensive VM type may not always result in optimal performance. Figure~\ref{fig:motivation_variance_time} shows the distribution of the execution time when running on the most expensive VM types  (namely c4.2xlarge, m4.2xlarge and r4.2xlarge). For instance, if we look at the distribution of execution times for c4.2xlarge, we observe that c4.2xlarge is the best VM for 50\% of the cases. This means for the other 50\% of the workloads; the most expensive VM type does not guarantee the lowest execution time. We observe similar behavior in Figure~\ref{fig:motivation_variance_cost}, where the least expensive VM, c4.large, does not ensure the lowest deployment cost.



\begin{figure}[t]
 \centering
 \subfigure[Execution time.]{
 \includegraphics[width=.22\textwidth]{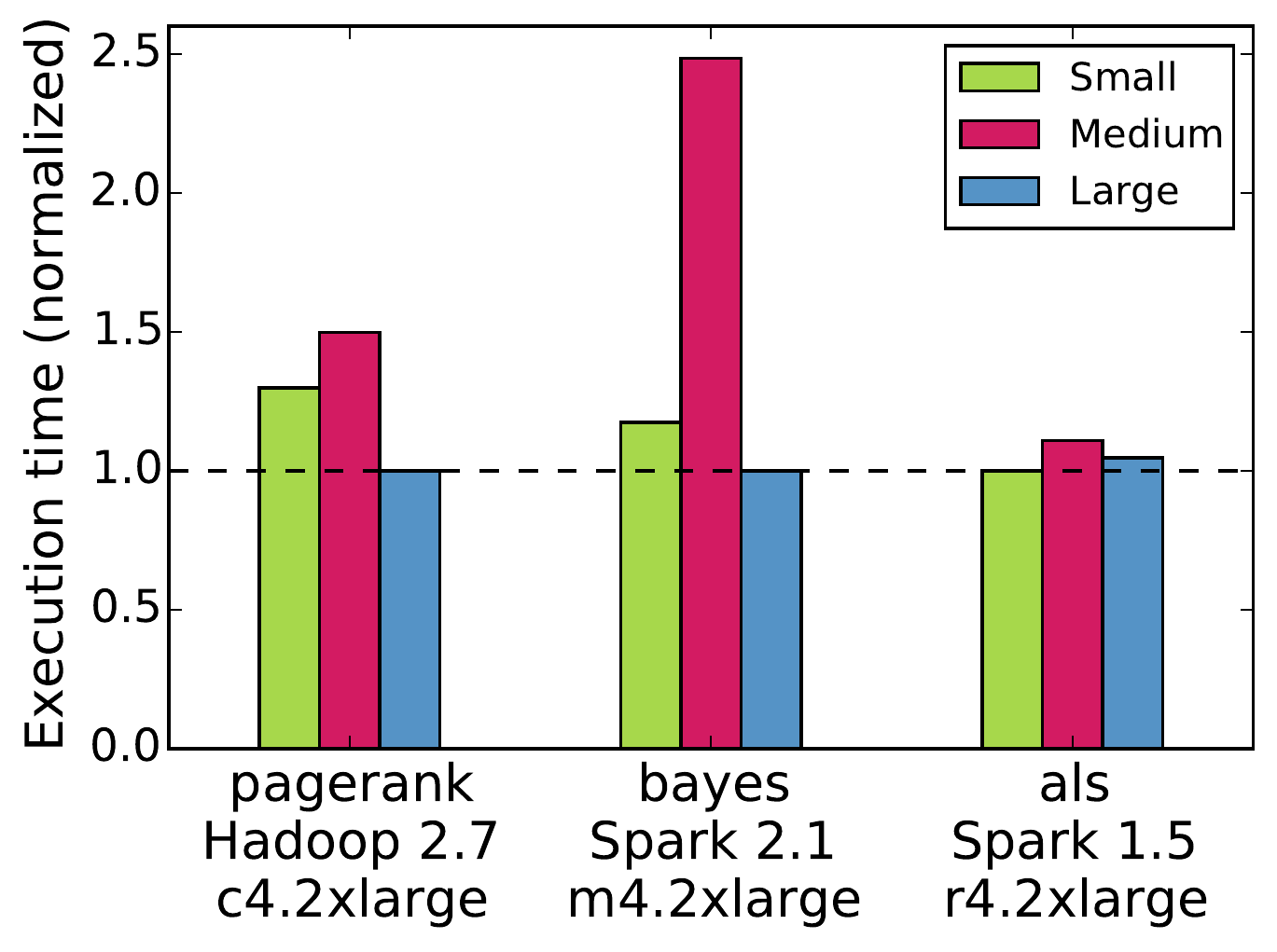}
 \label{fig:motivation_datasize_a}
 }
 \subfigure[Deployment cost.]{
 \includegraphics[width=.22\textwidth]{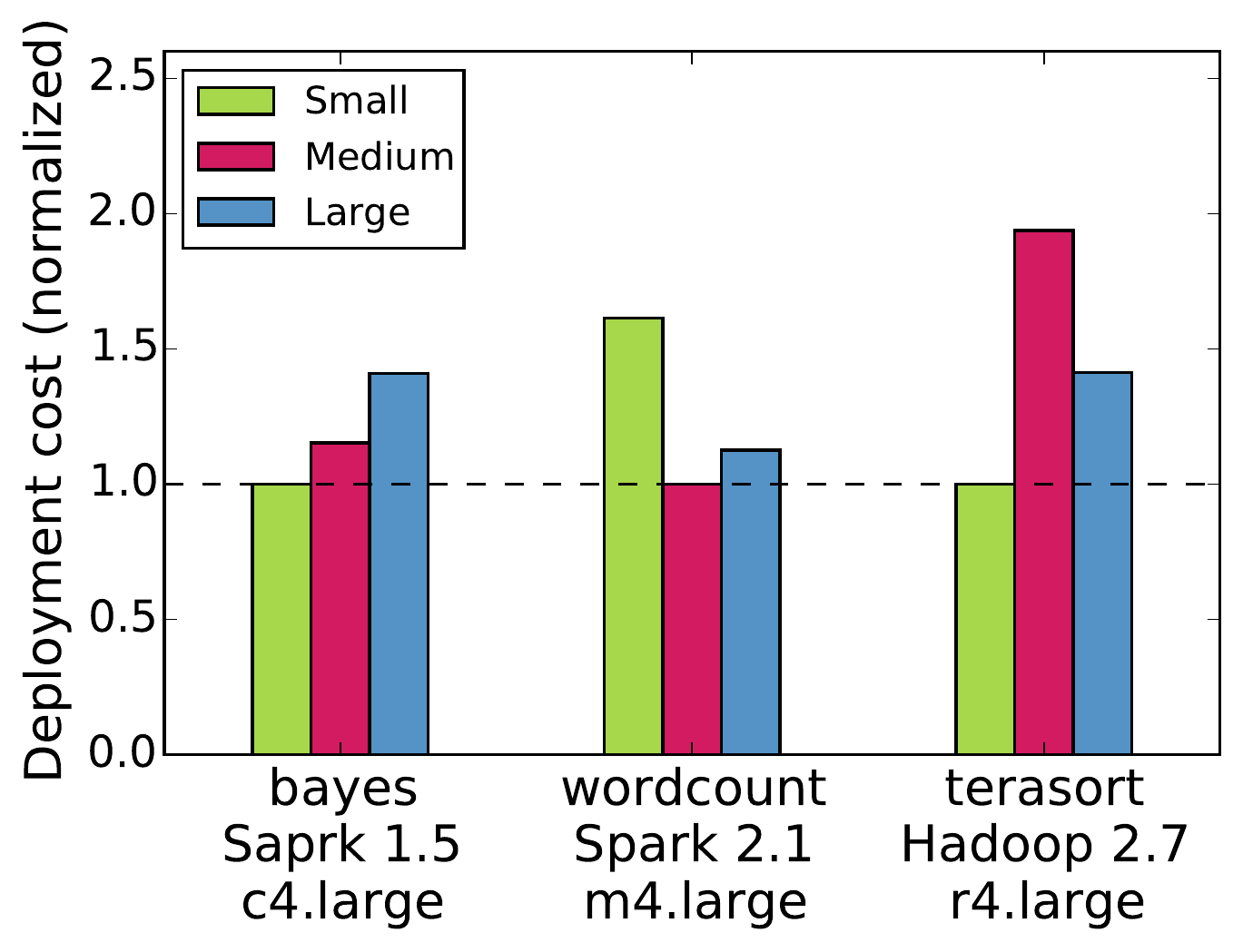}
 \label{fig:motivation_datasize_b}
 }
 \centering
 \caption{Running application with different input sizes result in very different performance.
 The best performing VM types for an application can change when the input size or parameters are changed.}
 \label{fig:motivation_datasize}
  \vspace*{-6mm}
\end{figure}

\noindent {\textbf{The same application with different input sizes favors different VM types:}}
Machine learning workloads are readily available
such as the machine learning library in Apache Spark and Python~\cite{scikit-learn}.
It is valid to assume that similar workloads would prefer the same VM type provided the user can accurately identify similar workloads.
Consequently, users can always reuse the best VM type for their workloads without testing further.
However, we found that this might not always be the case.
A workload with different input sizes or parameters performs very differently on different VMs~\cite{Venkataraman2016}.
Figure~\ref{fig:motivation_datasize} illustrates how the performance of an application varies with different input sizes.
For example, in Figure~\ref{fig:motivation_datasize_b}
\emph{m4.2xlarge} is the most cost-effective VM type for running the \textit{bayes} application with the \emph{small} input size.
However, the deployment cost increases by 40\% (is no longer the optimal VM) when the input size is \emph{large}.
A possible explanation is that a larger input size creates
a resource bottleneck on a smaller VM.
Hence, users need to be more careful
at selecting the best VM type even for similar workloads.

\noindent{\textbf{Cost creates a level playing field:}}
Finding a cost-effective VM type can be harder because
a slower VM can be competitive in deployment cost.
In \myfigure{\ref{fig:motivation_variance_time}},  \emph{c4.2xlarge} is the fastest VM type for over 50\% of the workloads (optimal execution time is 1.0).
However in Figure~\ref{fig:motivation_variance_cost}, when considering deployment cost, we observe that
\emph{c4.large} is likely to be a better choice, since it is optimal VM in over 50\% of the workloads.

\begin{figure}
    \centering
    \includegraphics[width=0.4\textwidth]{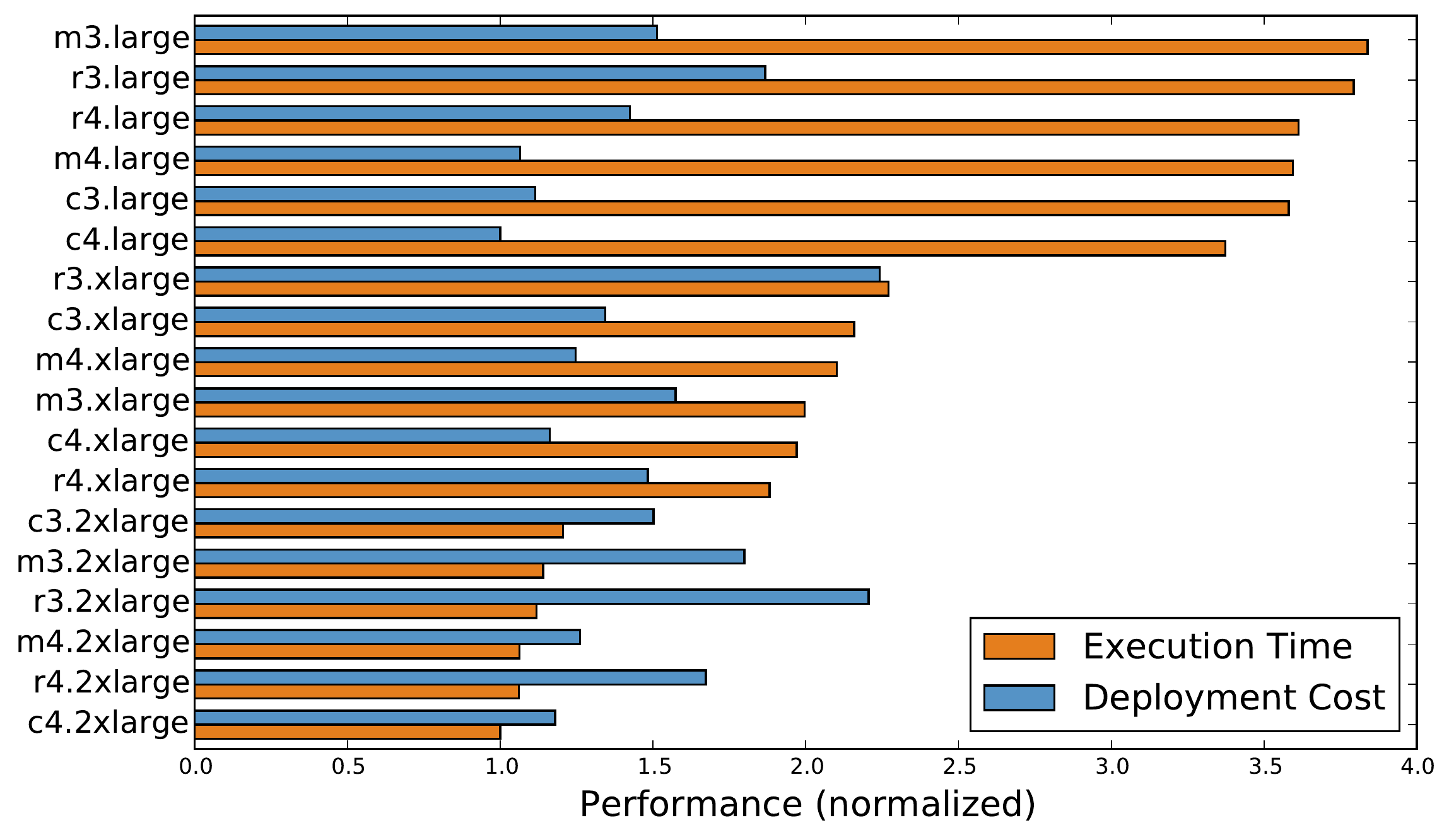}
    \caption{The performance of running the \emph{regression} workload on instances with different VM types.
    Introducing cost creates a \emph{level playing filed}, in which several inferior VM types in execution time are now competitive in deployment cost.
    This observation implies that searching for the most cost-effective configuration is harder than searching for the fastest configuration.}
    \label{fig:level_playing_field}
     \vspace*{-6mm}
\end{figure}

\myfigure{\ref{fig:level_playing_field}} presents the normalized execution time
and deployment cost of a workload (\emph{regression} on \emph{Spark 1.5}).
The figure demonstrates how execution time can be very different
while deployment cost is similar across all VM types. For example, \emph{m4.2xlarge} and \emph{r4.xlarge} are comparable to \emph{c4.2xlarge}.
When the difference between execution times of a workload in different VM types is large, choosing the best VM is easier because there is a clear winner.
Incorporating cost compresses the difference.
Therefore, searching for the most cost-effective VM type becomes more difficult because several inferior choices (in terms of execution time) are now competitive (in deployment cost).
In Section~\ref{sec:comparison}, we show why finding cost-effective VM type
is harder than execution time.

\subsection{Search-based optimization instead of complex models}
\label{sec:smbo}
One way to choose the best VM for a given workload is to build a complex prediction model from measurements as done in~\cite{Yadwadkar2017}.
However, this approach may encounter several obstacles.
First, the method assumes that data collection is free of noise.
However, this is not true in a cloud environment due to the sharing infrastructure, and therefore,
performance interference is unavoidable~\cite{Novakovic2013}. Second, a large amount of training data is required for building such complex models---which is not viable since each execution is expensive.
Moreover, even with the availability of data, performance predictability remains an issue.
For instance, \emph{PARIS} shows up to $50\%$ RMSE (Root Mean Squared Error) while predicting performance. 

Sequential model-based optimization (SMBO) iteratively measures solutions (VM types) to optimize for an objective (execution time or deployment cost)~\cite{Dewancker2015}.
SMBO is naturally applicable to finding the best VM. 
A typical SMBO algorithm is described in Algorithm~\ref{alg:smbo}. An SMBO algorithm requires 4 inputs namely, a cloud set up to run a workload ($f$), list of VM characteristics ($vm\in\mathit{VM}$) or instance space, an acquisition function ($S$), and a choice of surrogate model ($M$). SMBO starts with an initial sample of VMs (chosen randomly), which are then measured ($D$). Line 1. SMBO builds a surrogate or a machine learning model to estimate to predict workload performance.
This model is constructed using VM characteristics and the measured performance. Line 2.
A VM is selected based on the surrogate model along with a predefined acquisition function (see Section\ref{sec:BO}). Line 4. The selected VM ($x_i$) is then measured ($f$).
Line 5.
The VM ($x_i$) along with performance ($y_i$) is then added to the already measured VMs ($D=\{(vm_1, y_1), (vm_2, y_2)\}$). Line 6. This process terminates after a stopping criterion is reached.

We prefer SMBO because it is resistant to
the shortcomings in the complex-model building method, and
is suitable for optimizing any expensive black-box function.

\input{algorithm/smbo}

%% file: tables/dataset.tex
\begin{table}[t]
\centering

\caption{Applications evaluated in this paper. In total, there are 30 applications and 107 workloads measured on Hadoop 2.7, Spark 1.5 and Spark 2.1.}
\label{tab:dataset}
{\scriptsize
\begin{tabular}{@{}p{1.1cm}p{7.2cm}@{}}
\toprule
\textbf{Application} & \textbf{Description} \\ \midrule
\multicolumn{2}{l}{\noindent{\textbf{Micro Benchmark}}} \\
sort & Sorts text input data, generated by RandomTextWriter in Hadoop. \\
terasort & A standard Hadoop benchmark. Data is generated from TeraGen. \\
pagerank & The PageRank algorithm. Hyperlinks follow the Zipfian distribution. \\
wordcount & Counts the frequency of words that generated by RandomTextWriter.  This is a typical MapReduce job. \\ \midrule
\multicolumn{2}{l}{\textbf{OLAP}} \\
aggregation & Hive queries simulates OLAP-style queries as described in {[}15{]}. \\
join & Implement the join operation in Hive \\
scan & Implement the scan operation in Hive \\ \midrule
\multicolumn{2}{l}{\textbf{Statistics Function}} \\ 
chi-feature & Chi-square Feature Selection. \\
chi-gof & Chi-Square Goodness of Fit Test. \\
chi-mat & Chi-square Tests for identity matrix. \\
spearman & Compute Spearman's Correlation of two RDDs. \\
statistics & Generate column-wise summary statistics. \\
pearson & Compute the Pearson's correlation of two series of data. \\
svd & Singular Value Decomposition, a fundamental matrix operation for finding approximate solutions.\\
pca & Principal Component Analysis for dimension reduction. \\
word2vec & Generate distributed vector presentation of words according to distance. \\ \midrule
\multicolumn{2}{l}{\textbf{Machine Learning}} \\
classification & Implement the generalized linear classification model. \\
regression & Generalized Linear Regression Model. \\
als & The Alternating Least Squares algorithm, implemented in spark.mllib. It is a collaborative filtering algorithm used for product recommendation. \\
bayes & Implements the Naive Bayes algorithm for the multiclass classification problem. Input documents are generated from /usr/share/dict/linux.words.ords. \\
lr & A popular algorithm for the classification problem. \\
mm & Matrix multiplication with configurable row, column and block sizes.\\
d-tree & A greedy algorithm for classification and regression problems. \\
gb-tree & Gradient Boosted Tree, an ensemble learning method for classification and regression problems. \\
df & The Random Forest algorithm for classification and regression problems. \\
fp-growth & The FP-growth algorithm to mine frequent pattern in large-scale dataset. \\
gmm & Gaussian Mixture Model is a clustering algorithm that uses k Gaussian distributions to find the k clusters. \\
kmeans & K-means is a common clustering algorithm that finds k cluster centers. \\
lda & Latent Dirichlet allocation is a clustering algorithm that infers topics from a collection of text documents. \\
pic & Power iteration clustering is a scalable algorithm for clustering. \\ \bottomrule
\end{tabular}}
\end{table}

%% file: algorithm/smbo.tex
\begin{algorithm}[htbp]
{ \scriptsize
 \SetAlgoLined
 \DontPrintSemicolon
 \DontPrintSemicolon
 \KwIn{$f$, $\mathit{VM}$, $S$, $M$}
 \KwOut{The optimal configuration}
 $D$ := Initial sampling ($f$, $\mathit{VM}$) \;

 \For{$k$ in $|\mathit{VM} \notin D|$}{
   $p(y|vm, D)$ := fit a surrogate model (M, D) \;
   $vm_k$ := $arg max_{vm \in \mathit{VM}} S(vm, p(y|vm, D))$ \;
   $y_k$ := $f(vm_k)$ \;
   $D$ := $D \cup (vm_k, y_k)$ \;
 }
 \caption{Sequential Model-Based Optimization}
 \label{alg:smbo}
 }
\end{algorithm}

%% file: bo.tex
\section{Is Bayesian Optimization Fragile?}
\label{sec:bo}

In this section, we introduce Bayesian Optimization and
explain why BO can be fragile
in our problem setting.

\begin{figure}[t]
 \centering
 \subfigure[Minimizing execution time for \emph{als}]{
 \label{fig:cases_good}
 \vspace*{-2mm}
 \includegraphics[width=.35\textwidth]{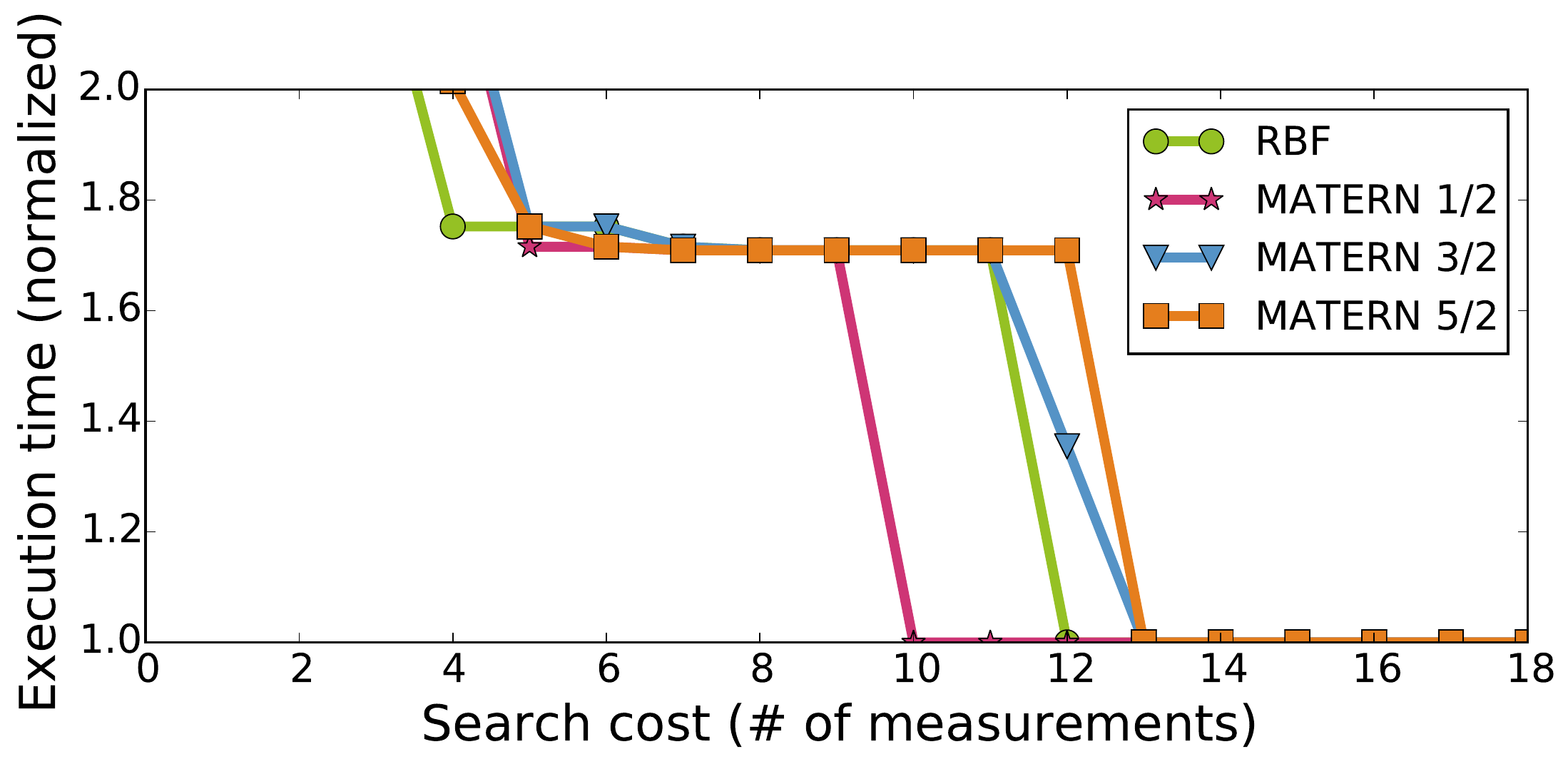}
 }
 \hfill
 \centering
 \subfigure[Minimizing deployment cost for \emph{bayes} ]{
 \label{fig:cases_bad}
 \vspace*{-2mm}
 \includegraphics[width=.35\textwidth]{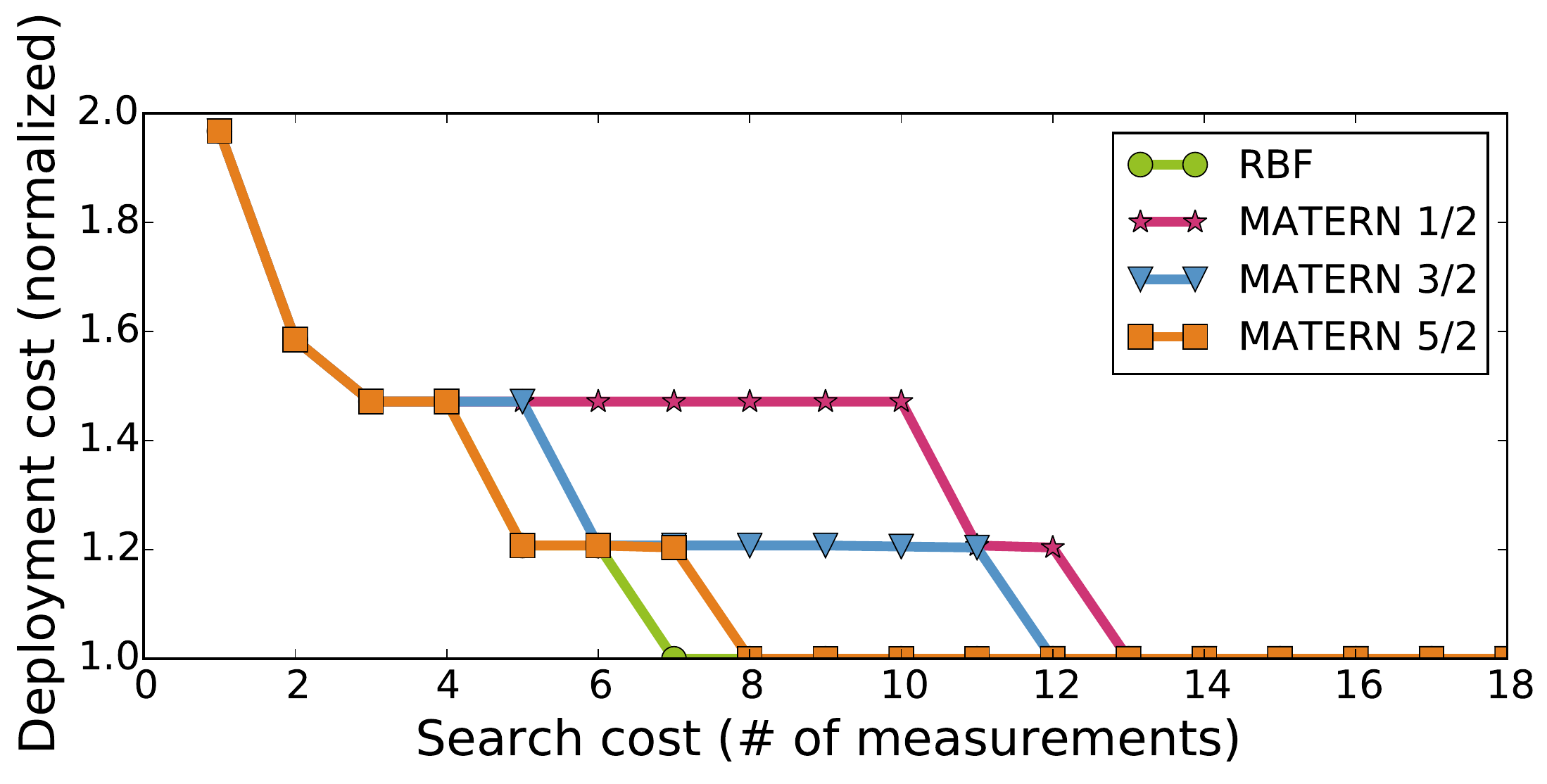}
 }
 \centering
 \caption{The number of actual measurements required to find
 the optimal VM type by Bayesian Optimization
 with different kernel functions.  Each kernel function is tested
 with 100 different sets of initial points uniformly selected. The line represents the median performance from 100 runs.}
 \label{fig:kernel_comparison}
 \vspace*{-6mm}
\end{figure}

\subsection{What is Bayesian Optimization?}\label{sec:BO}

BO follows the same formalism of sequential model-based optimization (SMBO) (as described in Section~\ref{sec:smbo}). Like SMBO, BO has two essential components namely a (probabilistic) regression model, and an acquisition function (Refer to~\cite{shahriari2016taking} for more details.)
BO has been used as a drop-in replacement to standard techniques such as
random search, grid search and manual tuning in numerous domains such as hyperparameter tuning and software performance optimization~\cite{Dewancker2015, Golovin2017, nair2017flash, zuluaga2016varepsilon}.
Recently, \emph{CherryPick} used BO to find the best VM for a specific workload~\cite{Alipourfard2017}.

In BO, Gaussian Process is
the standard probabilistic model used for building the surrogate model.
Gaussian Process is a distribution over objective functions specified by a
mean function and covariance function.
Once a surrogate model is trained, it can be used to estimate performance (of a workload) on the unmeasured VM. The surrogate model returns distribution of the estimated performance associated with the VM (mean and variance). The next VM to measure is determined by an acquisition function. Common acquisition functions are Probability of Improvement (PI),
Expected Improvement (EI), and Gaussian Process Upper Confidence Bound (GP-UCB)~\cite{Shahriari2016}.
Recently, the entropy search methods, backed by information theory, are promising alternatives~\cite{pmlr-v70-wang17e}.
In practice, EI is effective and used in \emph{CherryPick}.

An important component in Gaussian Process is the covariance kernel function, which is crucial for model effectiveness. Covariance kernel ensures that the prior, required for GP to be effective, is met. GP assumes smoothness, or in other words, the VMs which are closer to each other in instance space have similar performance.
This is particularly difficult in our problem setting, where a slight difference in the instance space and lead to significant differences in performance (cost or time). This goes to show that before using BO (with GP as a surrogate model), a practitioner needs to choose a kernel function to ensure smoothness in the instance space. Aforementioned could be particularly challenging and can affect the performance of BO. \emph{CherryPick} chooses the \emph{Mat\'ern 5/2} kernel function because it does not require strong smoothness, which are the cases for many real-world applications~\cite{Alipourfard2017,Yadwadkar2017}.

\subsection{How to choose the right covariance kernel function?}
\label{sec:kernel}
Since the choosing the covariance kernel function is critical, this section examines how different kernel function can affect the usefulness of BO.
We implement a BO (as prescribed by \emph{CherryPick}) to examine four different kernel functions:
(1) \emph{RBF}:  Radial Basis Function is a widely used kernel. However, RBF considers the effects of features on the covariance equally~\cite{Brochu2010}, which may not be realistic.
\noindent {Mat\'ern} kernel function is another family of covariance functions which incorporates a smoothness parameter such that it is flexible to model different objective functions. The smoothness parameter serves as similarity function that determines whether two samples are alike. 
The most commonly used smoothness parameters have three kinds namely, (2) \emph{Mat\'ern 1/2}, (3) \emph{Mat\'ern 3/2}, and (4) \emph{Mat\'ern 5/2}.

\myfigure{\ref{fig:kernel_comparison}} shows the number of actual measurements required to find the best VM for a given workload. In Figure~\ref{fig:cases_good}, shows how BO with \emph{Mat\'ern 1/2} kernel can find the optimal VM faster thereby reducing the search cost. However, in Figure~\ref{fig:cases_bad}, while trying to find a cost-effective VM,  BO with \emph{Mat\'ern 1/2} kernel performs the worst. The two particular examples we want to demonstrate how choosing the appropriate kernel function affects the performance of BO. In practice, choosing the right kernel function relies on engineering and automatic model selection~\cite{Brochu2010, Snoek2012, Dewancker2015, Shahriari2016}.

Our prior experience~\cite{Hsu2016} indicates that it is possible to have non-smooth performance outcome for a given workload on different VMs.
When a workload hits a resource bottleneck, \eg{memory or disk}, it can slow down greatly. This means that a workload might perform very differently on two VMs which are close to each other in the instance space. Therefore, we doubt that architecture parameters along are not sufficient to predict the performance of cloud applications~\cite{Yadwadkar2017, Hsu2016}.

\subsection{No one-size-fits-all initial points!}
\label{sec:init_points}

The choice of initial VMs also affects the effectiveness of BO. A common approach is a quasi-random method which uniformly selects very distinct VMs~\cite{Sobol1998}.
This method helps capture workload behavior, which can then be used to choose the next best VM to measure.
However, in practice, we have seen that BO is sensitive to initial points (VMs in our setting) and can exhibit large variances in their outcome.

To demonstrate the effect of initial VMs on the performance of BO, we choose three very different starting points, \ie{\emph{c4.xlarge}, \emph{m4.large} and \emph{r3.2xlarge}}, and then
run BO on all the 107 workloads. We observe that about 15\% applications do not find
the optimal configuration within six attempts (33\% of the instance space). 
We then redo the same experiment but choose different initial VMs. We find that
the BO can find the optimal configuration within six attempts.
This result demonstrates that the initial points dramatically affects the performance of BO.

Even though there exists a set of initial points that work well
on almost all applications,
the optimal initial VMs are subject to change because new VMs
are frequently added to the Cloud portfolios. Therefore, it is essential to design a search method that performs consistently with different initial points.

\subsection{Summary}

BO is a promising technique for finding the best VM for any workload.
However, our large-scale evaluation shows that a BO method can be \textit{fragile} or unstable.
Without proper design, it may lead to high search cost or a sub-optimal solution. This is because the effectiveness of BO is significantly affected by choice of the kernel function and the initial VMs (used to seed the BO). However, choosing the suitable kernel function requires
further analysis and in-depth study.
To sum up, BO can be fragile and requires extra care while making design choices. Our objective is to attempt to make BO less fragile by (i) augmenting BO with additional (low-level) information and (ii) use variants of BO - which are less sensitive.

%% file: approach.tex
\section{Low-Level Insight}
\label{sec:approach}

In this section, we introduce how to leverage low-level performance information to augment Bayesian Optimization.

\subsection{Choosing the Low-Level Metrics}

\begin{figure}
    \centering
    \includegraphics[width=0.45\textwidth]{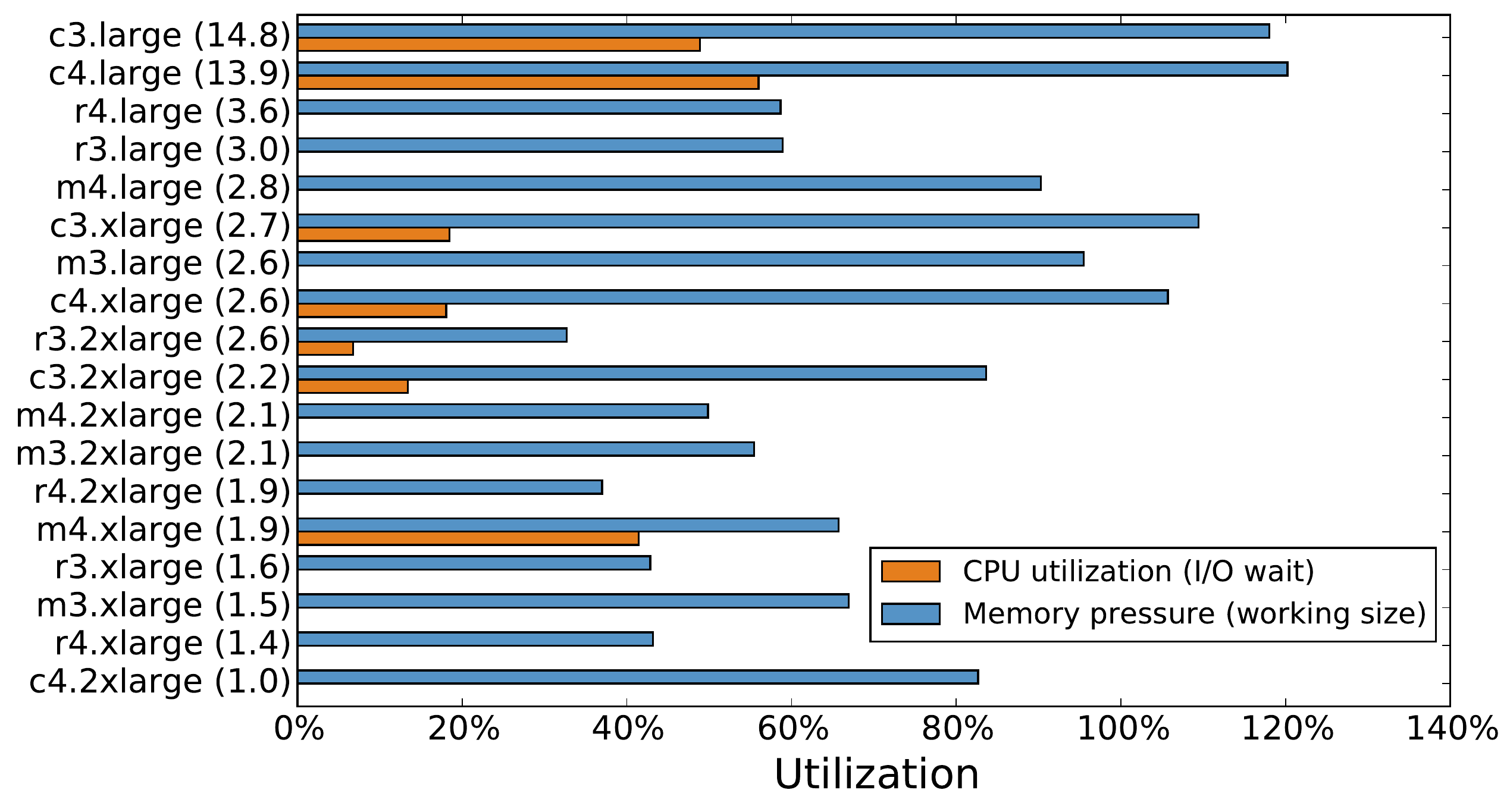}
    \caption{A memory bottleneck is identified by low-level performance information. The horizontal axis represents the resource utilization (\%), and the vertical axis represents the VM types. The numbers in the parenthesis are the normalized execution time where 1.0 represent the best VM type. The memory of a small VM type (c3.large) is not sufficient to run Logistic Regression, which leads to 14.8 times slower than the best VM type (c4.2xlarge). This behavior is captured by memory pressure and CPU utilization. 
}
    \label{fig:bottleneck}
\end{figure}

Prior work has shown how low-level performance metrics of workload are information, which is a good proxy for predicting performance~\cite{Novakovic2013, Hsu2016, Ousterhout2017, Yadwadkar2017}. For example, the \textit{memory commit size} represents the amount of memory required to handle current workload, and the \textit{CPU time waiting} on I/O indicates the workload type or a bottleneck on I/O. However, in practical settings, we need to analyze multiple metrics to better understand the factors which affect performance. System utilities on Linux, such as \emph{sysstat}, provides a comprehensive set of performance metrics~\cite{sysstat},
which are useful to characterize workloads and identify performance bottlenecks. In this work, we use these low-level metrics to augment a BO. The intuition behind this design choice stems from the fact that only instance space (published VM characteristics) is inadequate to fully characterize a VM.
In \myfigure{\ref{fig:bottleneck}},
we show an example to use the low-level information to
to identify the memory bottleneck while running Logistic Regression.

Since we focus on recurring jobs, we should use metrics that can capture the workload progress and identify resource bottleneck.
The selection of low-level metrics depends heavily on workloads.
If possible, we should use a comprehensive set of metrics.
However, a large number of features can lead to the over-fitting problem in building predictive models.
This is known as the curse of dimensionality~\cite{domingos2012few}.
In this work, we find the following low-level metrics are effective.
Automatic feature selection can help address this problem~\cite{guyon2003introduction, Hsu2016} but will require further studies.

\begin{itemize}
    \item \textbf{Workload progress}: CPU utilization on user time, I/O wait time, and the number of tasks in the task list.
    \item \textbf{Memory pressure}: \% of commits in memory.
    \item \textbf{I/O pressure}: disk utilization and wait time (disk).
\end{itemize}

\subsection{Low-Level Augmented Bayesian Optimization}

Leveraging low-level information in BO requires novel modeling methods because the given workload is yet to be executed on the candidate VM. Our approach, instead, predicts the performance (cost or time) based on the VM characteristics and observed low-level metrics of the VM that is already measured. This is similar to the reasoning technique used in practical settings by experts and the table based models~\cite{Anderson2001}. Experts choose to interpolate or extrapolate the workload performance using not only characteristics of VM but also the low-level performance information.

We make the following design choices to modify Naive BO to integrate low-level performance information into BO:

\noindent\textbf{Augmented Instance Space}: Instead of using only VM characteristics ($\mathit{VM}_i$),
as an input to the surrogate model we also use low-level metrics ($L_i$) collected from running the workload on  ($\mathit{VM}_i$). These constitute the \textit{independent variables}. Similar to Naive BO, the performance of the workload is used as the \textit{dependent variable}. \textit{Decision to use low-level information allows BO to  make more informed search.}

\noindent\textbf{Surrogate Model}: Instead of using Gaussian Process as the surrogate model, we choose a tree-based ensemble method Extra-Trees algorithm for building the surrogate model. The tree-based learning method is effective
to capture complex performance behavior
~\cite{Wang2004, Yin2006, Noorshams2013,Hsu2016, Yadwadkar2017}. We choose not to use Gaussian Process in Bayesian Optimization because determining the right kernel function (as discussed in Section~\ref{sec:kernel}) requires careful evaluation, which is not practical for supporting diverse workloads. \textit{This design choice lets us side-step one of the reasons for the fragility of Naive BO}.

\noindent\textbf{Acquisition Function}: We replace \textit{Expected Improvement} (EI) with \textit{Prediction Delta} as the acquisition function. 
Prediction delta can select a VM type with highest estimated performance (least execution time or cheapest deployment cost). Prediction Delta can also be used as a stopping criterion---terminate the search process if there exists no better VM type. We do not use Expected Improvement as our acquisition function because it is not useful when the kernel function cannot estimate the black-box function. \textit{This design choice for an acquisition function which does not require a suitable kernel function.}

\noindent\textbf{Surrogate Model Update}: When updating the surrogate model upon a new observation for workload (w) ($\mathit{VM}_{i,w}$, $L_{i,w}$ and $y_{i,w}$), we generate multiple pairs of input ($\mathit{VM}_{j,w}$, $\mathit{VM}_{i,w}$), where $i\neq j$ with
low-level information ($L_{j,w}$), where $j$ represent the source VM---which has been measured and $i$ represents the destination VM---which is yet to be measured.
This surrogate model answers \quotes{what is the predicted performance of $\mathit{VM}_{i,w}$ given the low-level performance information observed on a particular VM ($\mathit{VM}_{j,w}$)}. For example, if we have measured the performance of workload (w) in 3 VMs ($VM_{1,w}, VM_{2,w}, VM_{3,w}$), the number of independent values for which the performance needs to be estimated would be $3 \times (18-3)$. It should be noted that to estimate the performance of a workload in a VM (say $\mathit{VM}_{15,w}$), requires considering $\mathit{VM}_{1,w}\rightarrow\mathit{VM}_{15,w}$,  $\mathit{VM}_{2,w}\rightarrow\mathit{VM}_{15,w}$, and $\mathit{VM}_{3,w}\rightarrow\mathit{VM}_{15,w}$.  Since multiple pairs exist, we average the estimated performance. \textit{This design choice helps us update the surrogate model even when the low-level information of destination VM is not available.}

\noindent Algorithm~\ref{alg:abo} illustrates the Augmented BO.
\input{algorithm/arrow}

%% file: algorithm/arrow.tex
\begin{algorithm}[htbp]
{
\scriptsize
 \SetAlgoLined
 \DontPrintSemicolon
 \DontPrintSemicolon
 \KwIn{$f$, $\mathit{VM}$, $S$, $M$}
 \KwOut{The optimal configuration}
 $D$ := Initial sampling ($f$, $\mathit{VM}$) \;

 \For{$k$ in $|\mathit{VM} \notin D|$}{
   $p(y|vm, D, L)$ := FitLowLevelModel(M, D, L) \;
   $vm_k$ := $arg max_{vm \in \mathit{VM}} S(x, p(y|vm, L, D))$ \;
   $y_k, L_k$ := $f(\mathit{vm}_k)$ \;
   $D$ := $D \cup (\mathit{vm}_k, y_k, L_k)$ \;
 }
 \caption{Low-Level Augmented Bayesian Optimization (similar to Algorithm~\ref{alg:smbo})}
 \label{alg:abo}
 }
\end{algorithm}

%% file: evaluation.tex
\begin{figure*}[t]
 \centering
 \subfigure[Optimizing running time]{
 \label{fig:overall_time}
 \includegraphics[width=.45\textwidth]{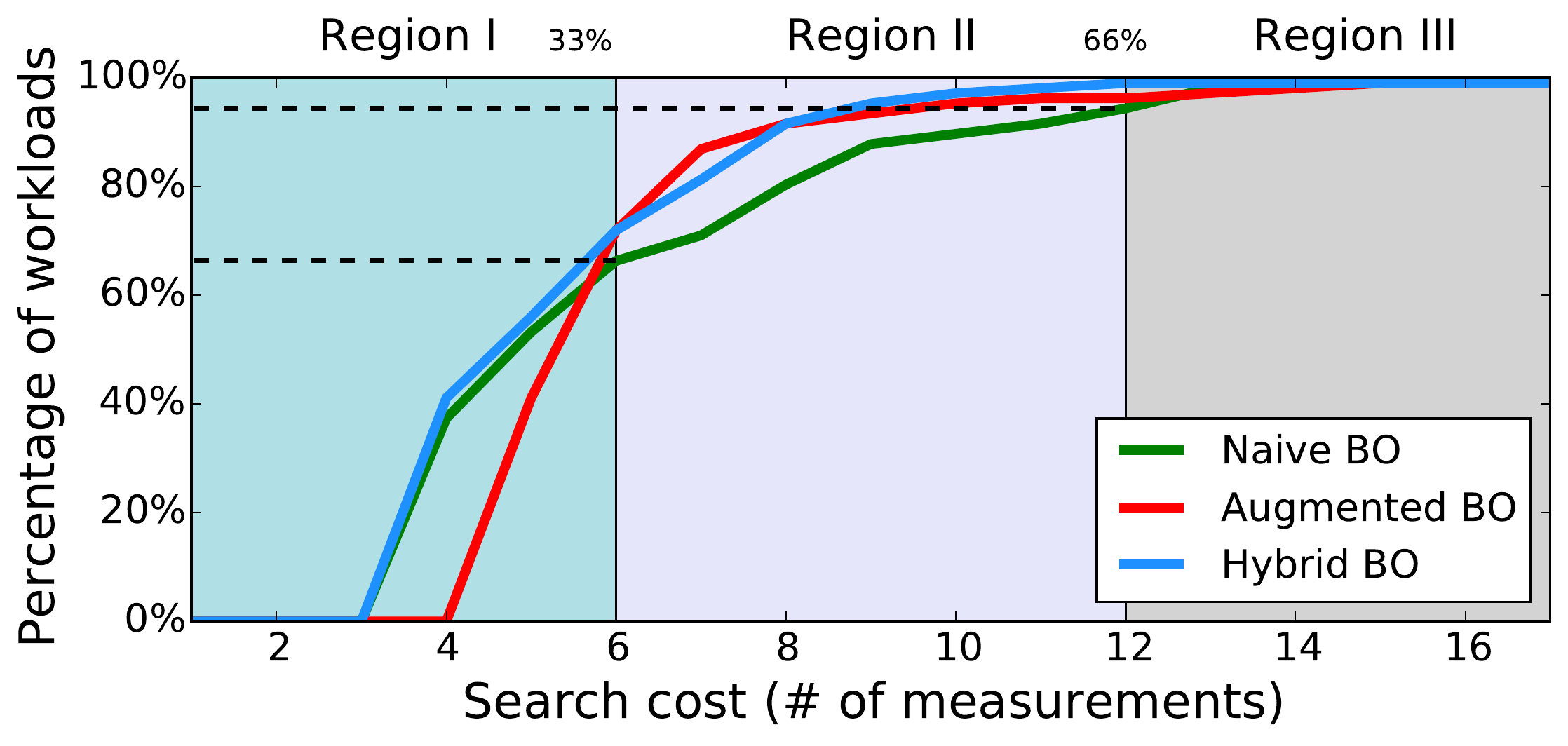}
 }
 \subfigure[Optimizing running cost]{
 \label{fig:overall_cost}
 \includegraphics[width=.45\textwidth]{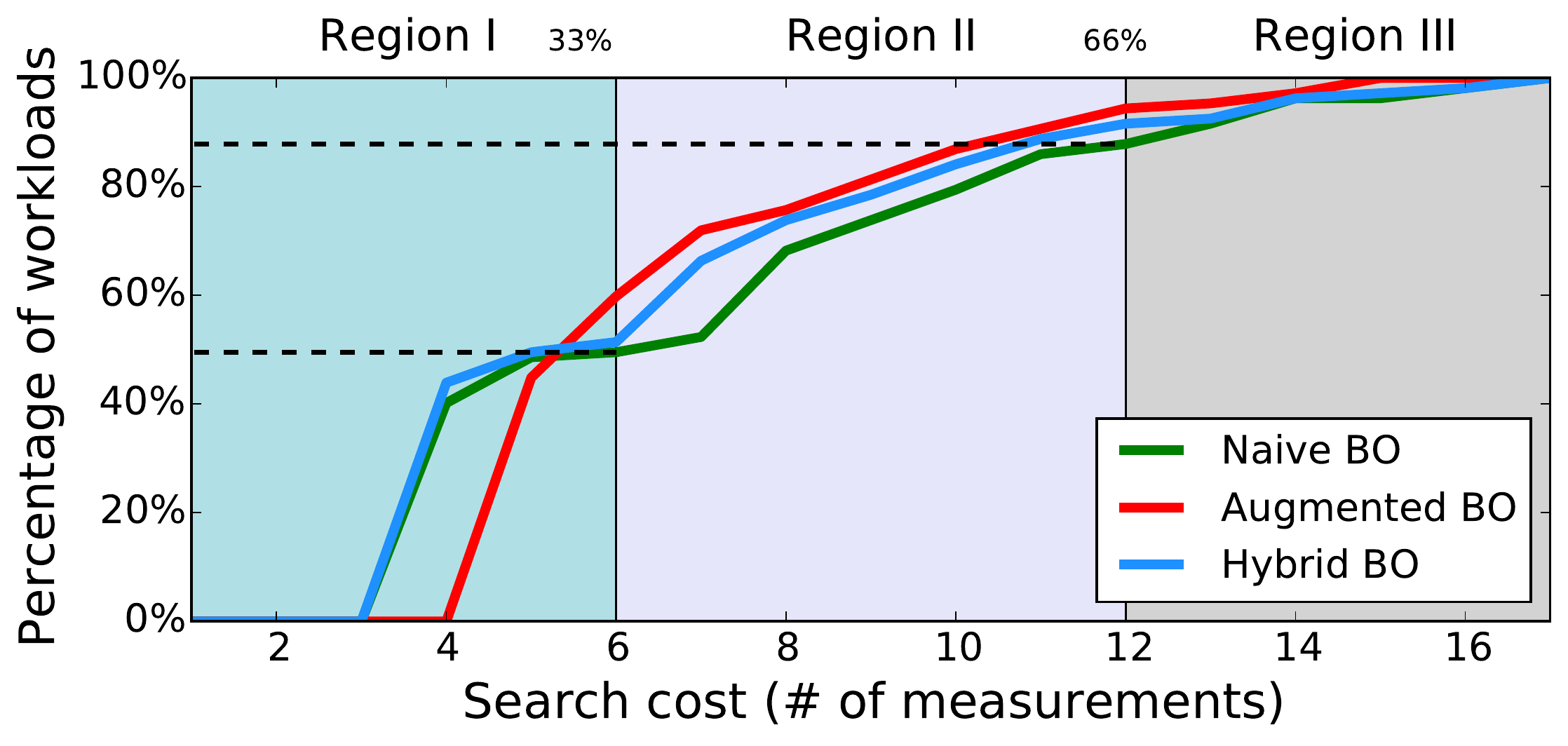}
 }
 \centering
 \caption{Search cost of finding the optimal VM type across the 107 workloads. The \emph{y-axis} represents the cumulative percentages of workloads.
 In \emph{Region I}, although Augmented BO does not find the optimal VM type at the fourth step, it does find a very near optimal solution with only 4\% difference.
 Section~\ref{sec:practice} provides further details.}
 \label{fig:overall}
 \vspace*{-6mm}
\end{figure*}

\section{Evaluation}
\label{sec:evaluation}

This section describes our experimental setting and evaluation method to compare Augmented BO with Naive BO.
\subsection{Experimental Method}
\label{sec:experiment}

\noindent {\textbf{Workload}}:
For evaluation, we use Apache Hadoop (v2.7) and Spark (v2.1 and v1.5), which are
popular systems for many big data and machine learning applications. We choose distinct workloads from \emph{HiBench} and \emph{spark-perf}, as listed in \mytable{\ref{tab:dataset}}.
\emph{HiBench} is a big data benchmark suite for Apache Hadoop and Spark \cite{hibench}. It was designed to test batch processing jobs and streaming workloads. Similarly, \emph{spark-perf} is a performance testing suite for Spark \cite{sparkperf}.
The testing suite provides a wide range of workloads including
supervised learning such as regression and classification modeling,
unsupervised learning such as K-Means clustering,
and statistical tools such as correlation analysis, and Principal Component Analysis (PCA).
We run 107 workloads to test their performance on 18 VM types.
During the execution of the workload, a \emph{sysstat} demon is run in the background to collect low-level performance information~\cite{sysstat}. We do not find any signification overhead to collect low-level information.

\noindent {\textbf{Cloud Configurations}}:
We measure the performance on six VM families (available on AWS)
\{\emph{c3}, \emph{c4}, \emph{m3}, \emph{m4}, \emph{r3} and \emph{r4}\}, and
three VM sizes \{\emph{large}, \emph{xlarge} and \emph{2xlarge}\}.\footnote{The latest generation has been upgraded
from \emph{c4} to \emph{c5} for the compute-optimized VM and from \emph{m4} to \emph{m5} for the general-purpose VM
after we completed our data collection.}
The VM size represents the core count.
For example, \emph{c4.large} has two cores,
\emph{c4.xlarge} has four cores and \emph{c4.2xlarge} has eight cores.

\noindent {\textbf{Encode Cloud Configurations}}:
We choose four VM characteristics namely CPU types, core count, average RAM per core, and the bandwidth to Elastic Block Storage (EBS). We encode the four features with numerical values into $\vec{\mathit{VM}}$. The CPU types are encoded from one to six in order, and for the core count, we use their actual values \{2, 4, 8\}.
Similarly, the RAM size per core is \{2, 4, 8\}. Last, the bandwidth to EBS has three classes for different VM types encoded as \{1, 2, 3\}.

%

\subsection{Comparison}
\label{sec:comparison}
This section evaluates Naive BO and Augmented BO, on the 107 workloads with randomly selected initial VMs. The above process is repeated 100 times to account for randomness.
Here we minimize execution time and deployment cost individually. 
\myfigure{\ref{fig:overall}} presents the overall result. 

\begin{figure*}[t]
 \centering
 \subfigure[PageRank on Hadoop 2.7]{
 \label{fig:convergence_time_1}
 \includegraphics[width=.3\textwidth]{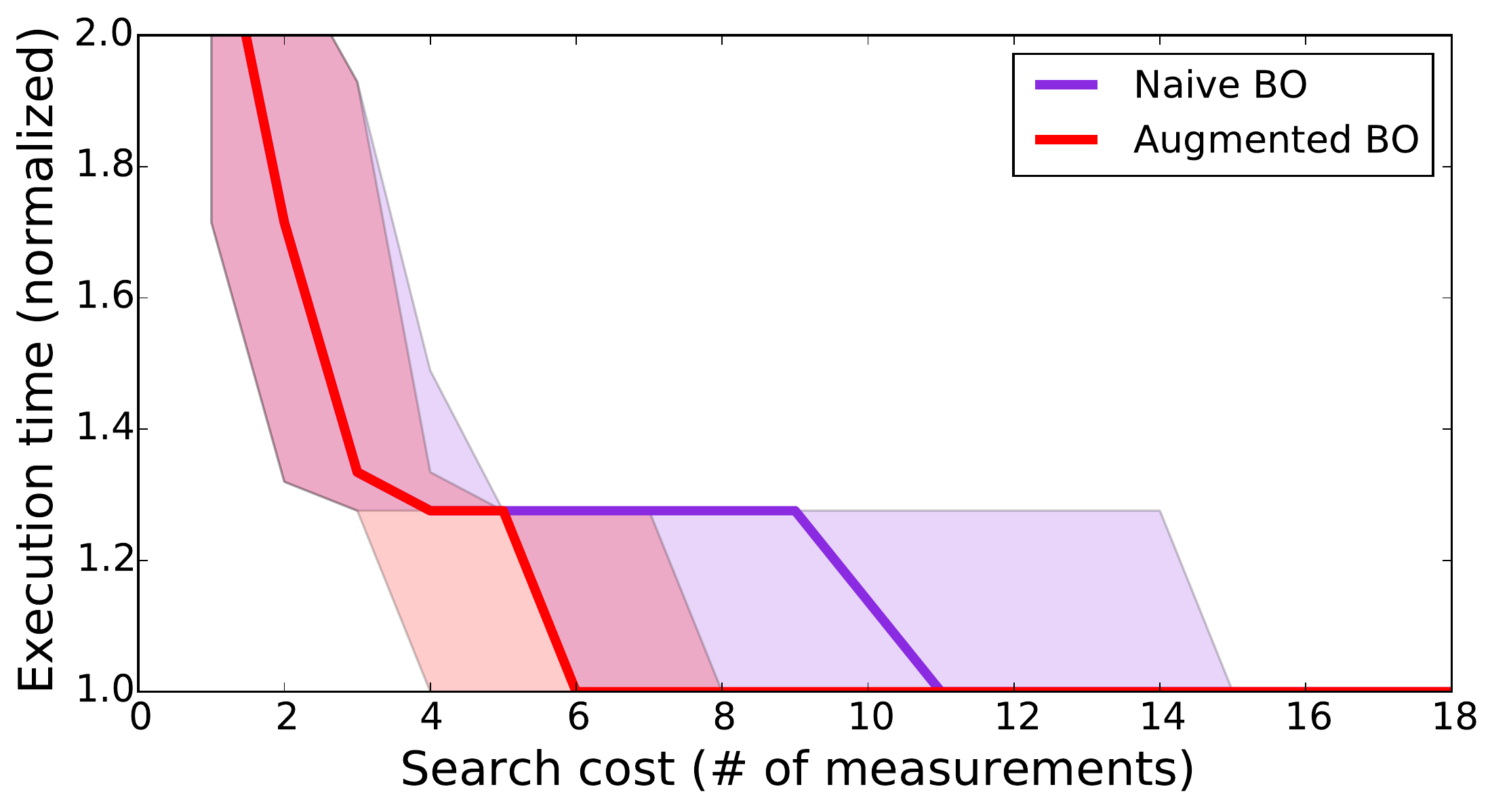}
 }
 \subfigure[Alternating Least Squares on Spark 2.1]{
 \label{fig:convergence_time_2}
 \includegraphics[width=.3\textwidth]{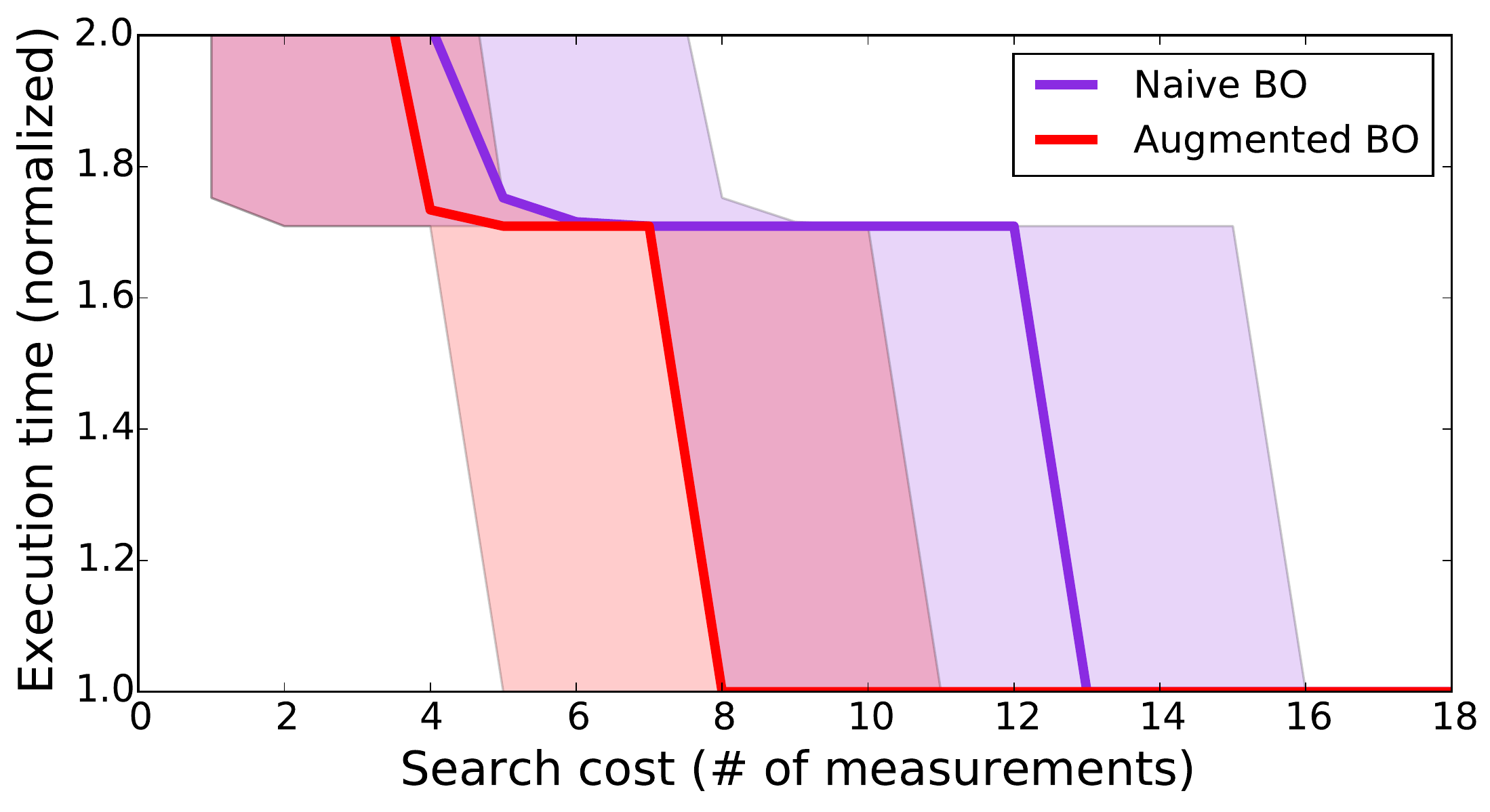}
 }
 \subfigure[Logistic Regression on Spark 1.5]{
 \label{fig:convergence_cost_2}
 \includegraphics[width=.3\textwidth]{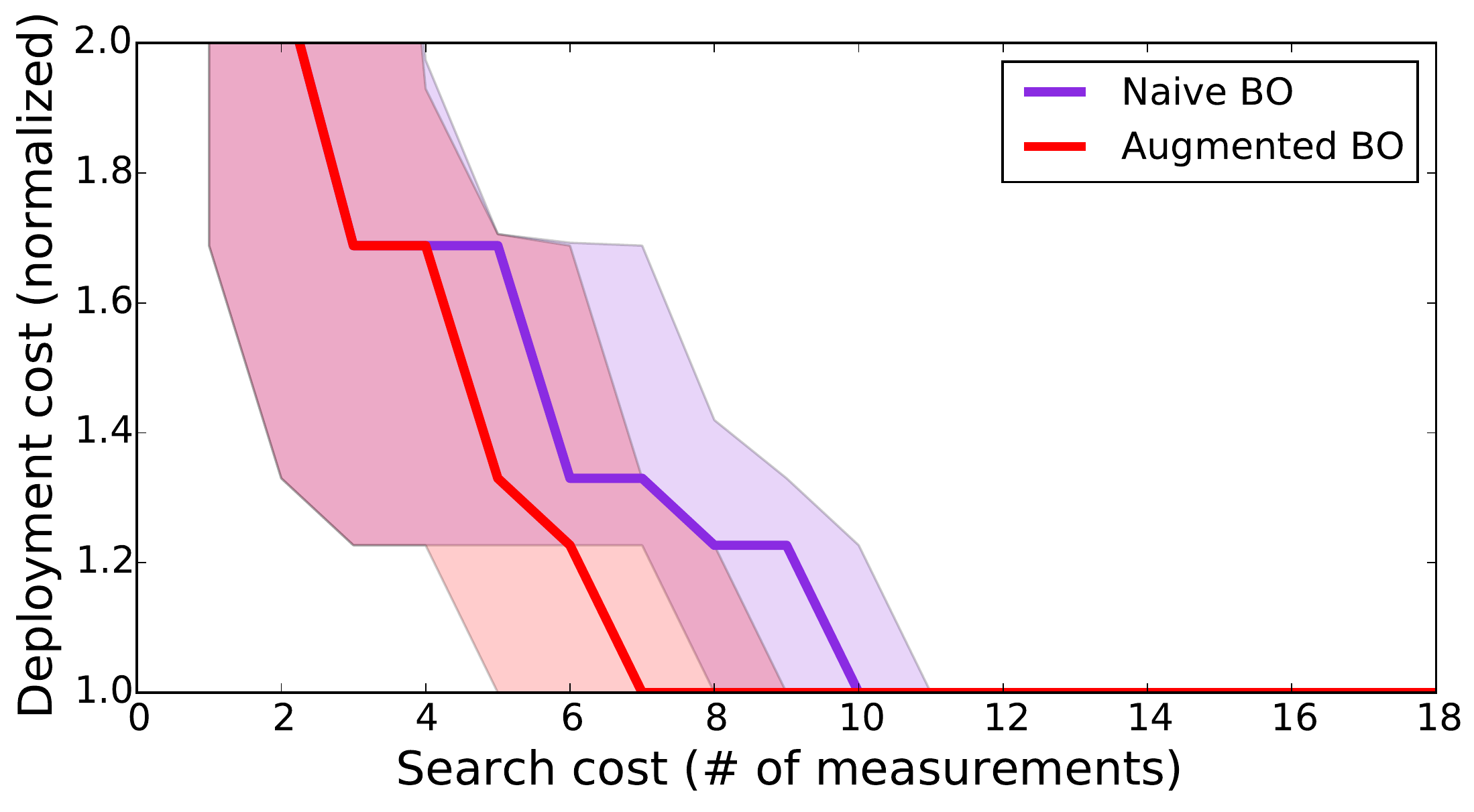}
 }
 \centering
 \caption{
Examples of searching for the best VM. The objective is to find the fastest VM in subfigures (a, b) and the most cost-effective VM in subfigure (c).  Both the BO methods stops after they find the optimal VM type (normalized to 1.0).  The line represents the median value of the execution time over 100 repeats. Each repeat used different initial points to seed BO. The shaded region represents the IQR or Interquartile range is the difference between $3^{rd}$ and $1^{st}$ quartile. A high value (larger area) of IQR indicates high variance.}
 \label{fig:convergence_time}
 \vspace*{-6mm}
\end{figure*}

\vspace{0.16cm}
\noindent {\textbf{RQ1: Can Augmented BO find optimal VMs?}}
\vspace{0.16cm}

\noindent Figure~\ref{fig:overall_time} shows the percentage of workloads, where Naive BO and Augmented BO found the optimal VM. The horizontal axis represents the search cost (in terms of \# of measurements), and the vertical axis represents the percentage of the workloads. In this figure the \textcolor{green}{green} line represent the Naive BO and the \textcolor{red}{red} line represent the Augmented BO. The Naive BO can find the optimal solution for 60\% of the workloads by searching 33\% of the search space (Region I). The performance of Augmented BO is similar to naive BO. However, Augmented BO has a slow start problem but becomes effective eventually. In Region II, Augmented BO is a clear winner as it can find optimal VMs for 96\% of the workloads within ten measurements. At the same time, Naive BO can only find 80\% of the workload.

We claim that the performance of Augmented BO is better than Naive BO, for regions I and II (at step 6 and 12). We also observe an interesting phenomenon---Augmented BO is outperformed by Naive BO in initial four steps. The one-step difference can be attributed to the over-fitting problem caused by
the larger training features (both high-level and low-level information) in Augmented BO. This is a challenge of leveraging low-level information (for future work).

While looking at the performance of VMs selected by Augmented BO, we observe that the best VM found by Augmented BO is only 4\% away from the optimal VM. In practice, this difference can be easily ignored (refer to
Section~\ref{sec:practice}). Furthermore, with the growing instance space,
this difference (though we believe is little) can be amortized because
the search cost will also increase.

A possible workaround to this problem can be to create a Hybrid BO (shown in \textcolor{blue}{blue})---which combines the best of the two methods.
\myfigure{\ref{fig:overall_time}} shows that \emph{Hybrid BO}
outperforms Naive BO for in all cases.
However, we choose not to focus on the hybrid method in this paper because of space constraints. Another reason why we do not discuss Hybrid BO because our primary objective is to identify the fragility of Naive BO and the advantages of leveraging low-level information.\footnote{Please refer \url{https://goo.gl/Yo5Gv3} for more details about Hybrid BO.}

\vspace{0.16cm}
\noindent {\textbf{RQ2: Can Augmented BO minimize cost?}}
\vspace{0.16cm}

\noindent To answer the question if Augmented BO can minimize the deployment cost, it is essential to demonstrate that Augmented BO can find optimal VM faster than Naive BO (lower search cost). In Figure~\ref{fig:overall_cost}, we observe that the \textit{minimizing deployment cost is more difficult} i.e., both methods require more search cost to reach the optimal solution. We observe that Naive BO can find the best VM with six attempts
for only 50\% applications while Augmented BO increases this probability to 60\%. We also see a clear win for Augmented BO as it can find best VM which minimizes the deployment cost after measuring five measurements. However, we see that Augmented suffers from a slow start, which is similar to Figure~\ref{fig:overall_time} and  Hybrid BO (shown in \textcolor{blue}{blue}) is the workaround.

\begin{figure*}[t]
 \centering
 \subfigure[Region I]{
 \label{fig:stopping_criteria_comparison_good}
 \includegraphics[width=.3\textwidth]{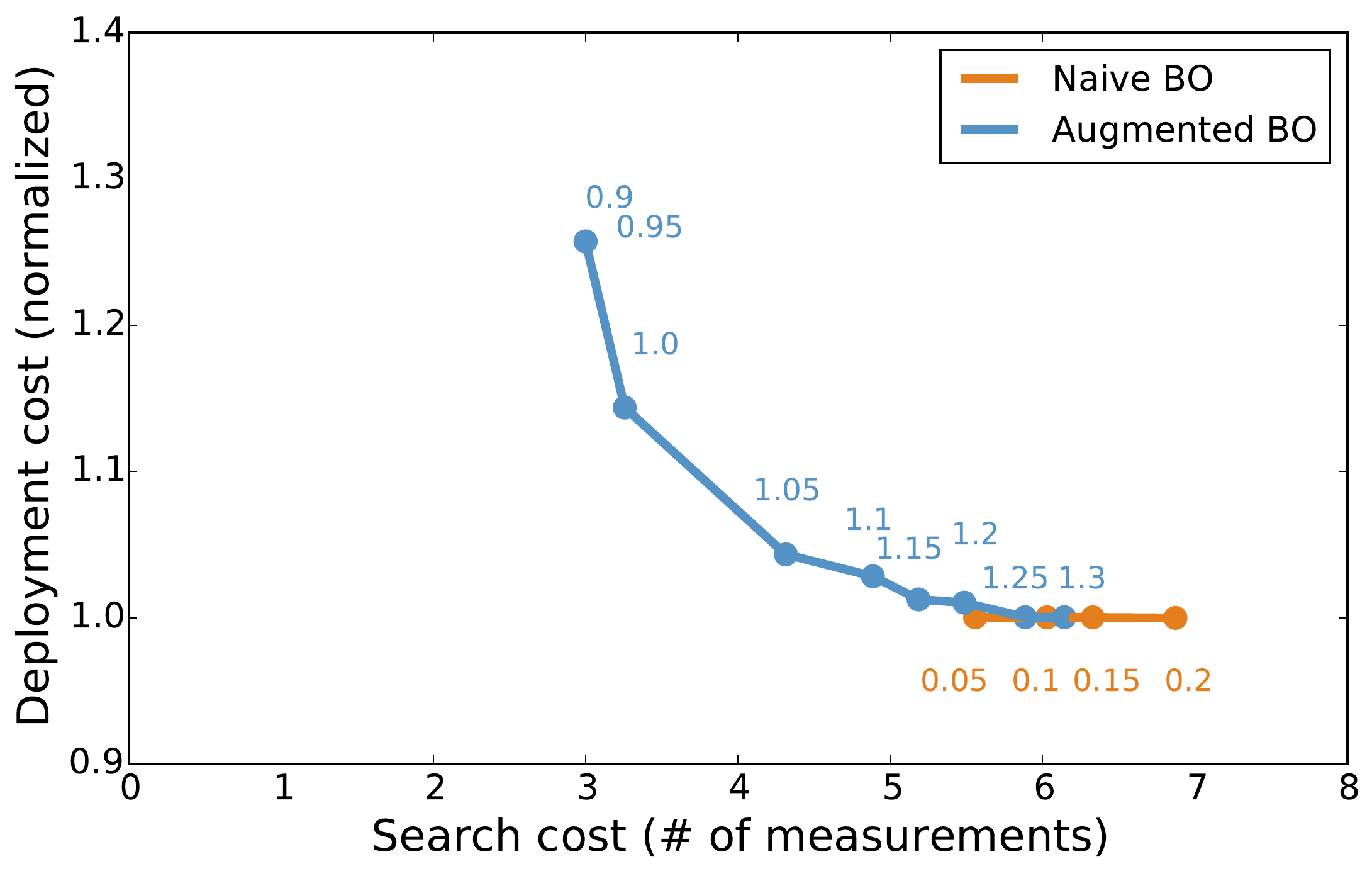}
 }
 \subfigure[Region II]{
 \label{fig:stopping_criteria_comparison_bad}
 \includegraphics[width=.3\textwidth]{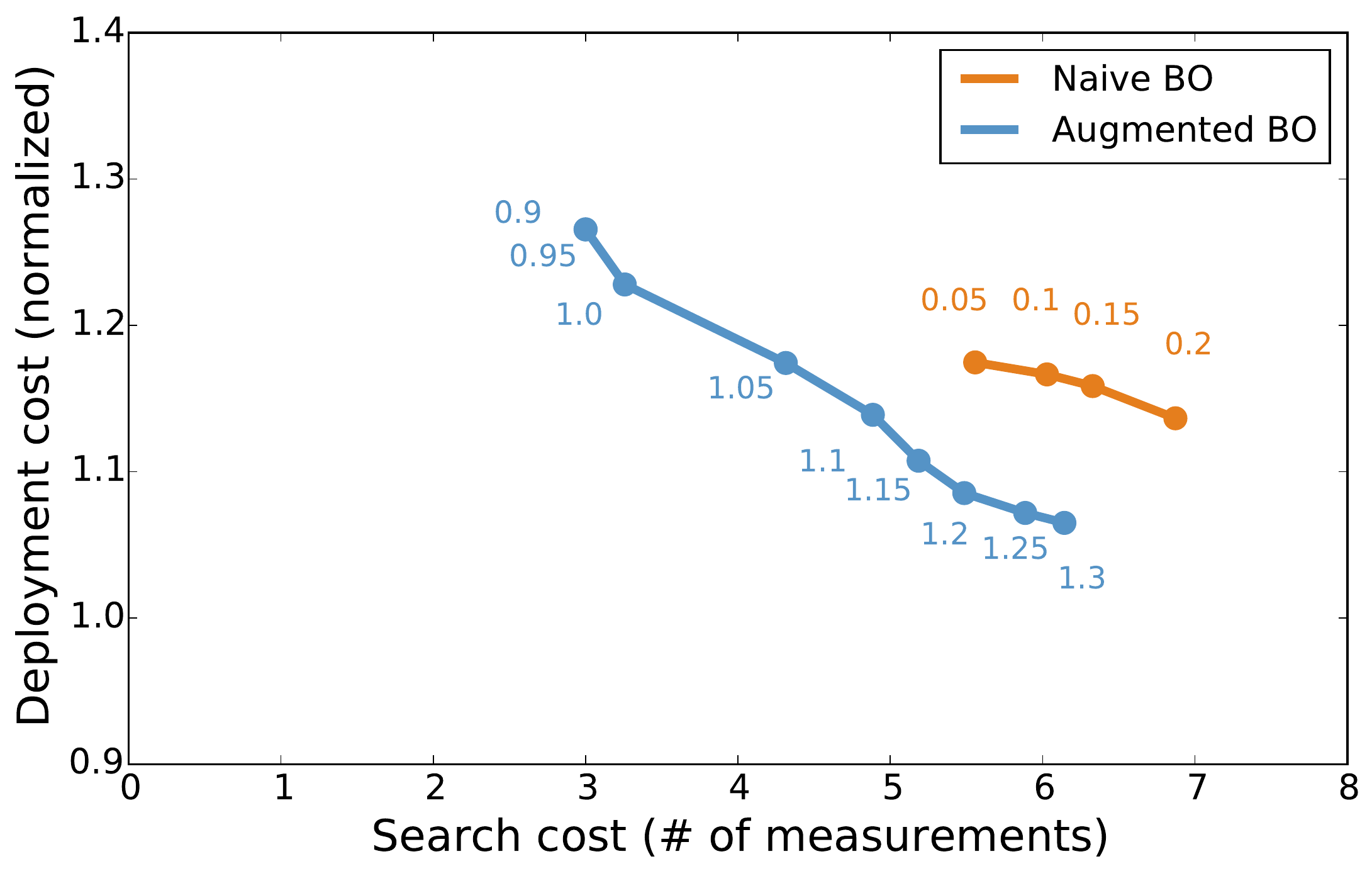}
 }
 \subfigure[Region III]{
 \label{fig:stopping_criteria_comparison_problematic}
 \includegraphics[width=.3\textwidth]{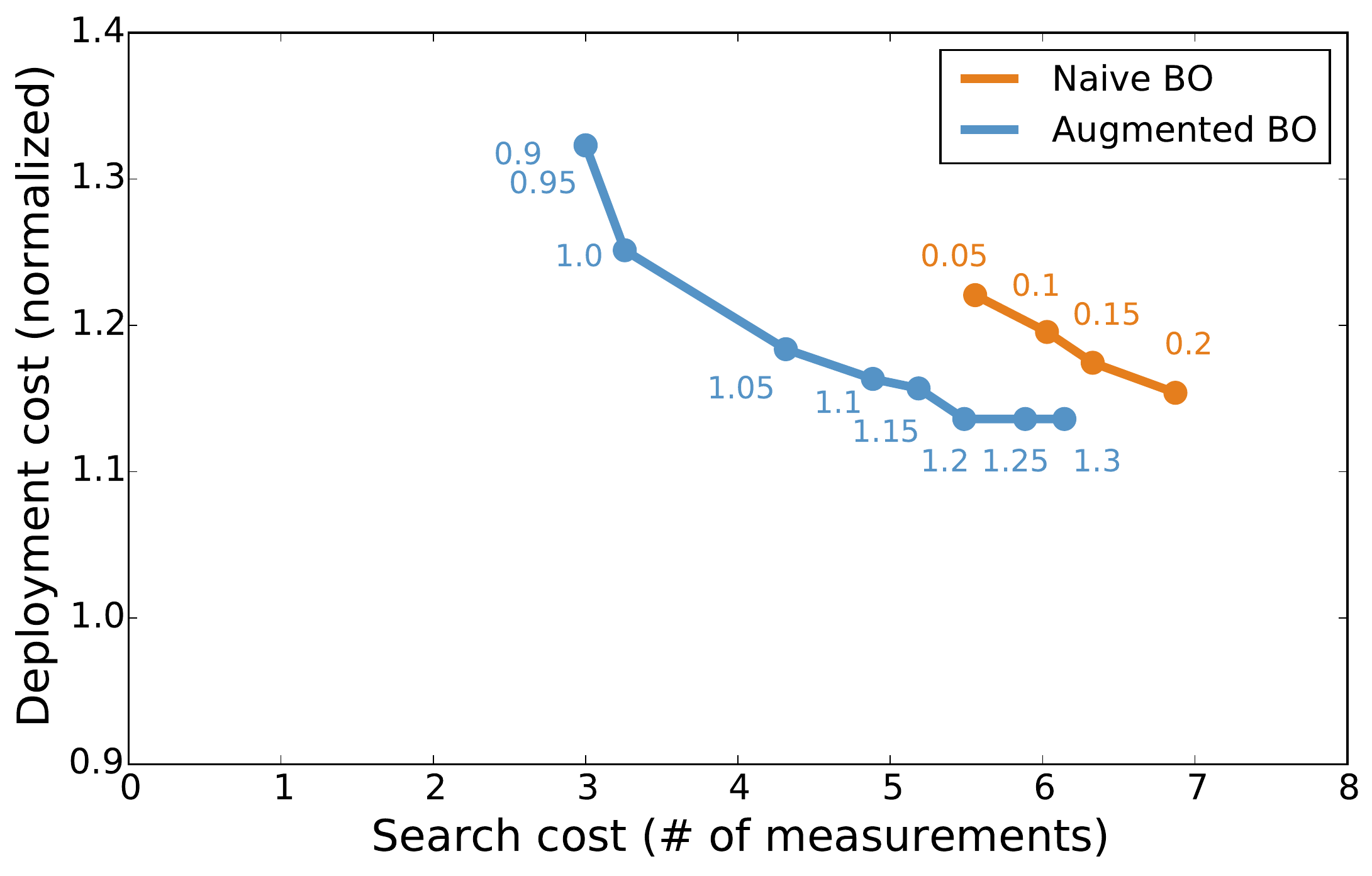}
 }
 \centering
 \caption{Comparison between effectiveness of search with different stopping criteria.  There is a trade-off between search cost and deployment cost. In \emph{Region I}, Augmented BO is comparable with Naive BO in terms of deployment cost but can greatly reduce search cost at the expense of slight increase in deploymwnr cost. For \emph{Region II} and \emph{Region III}, Augmented BO outperform Naive BO for both search cost and deployment cost.}
 \label{fig:stopping_criteria_comparison}
  \vspace*{-6mm}
\end{figure*}

\vspace{0.16cm}
\noindent {\textbf{RQ3: Is Augmented BO fragile?}}
\vspace{0.16cm}

\noindent In finding the best VM, Naive BO fails
in 36\% (minimizing time) and 50\% (minimizing cost) of the workloads
after measuring the performance of the six VMs (Region I).
Augmented BO alleviates this problem, and this can be observed by a
up to 20\% increase in the number of workloads for which Augment BO found the optimal VM (step 7 in~\myfigure{\ref{fig:overall_cost}}).

Stability is another important aspect of Augmented BO. As discussed in Section~\ref{sec:init_points},
initial points are critical to the performance of BO---different initial VMs can lead to very different results (performance and search cost) or high variances in results. \myfigure{\ref{fig:convergence_time}} compares the search cost and the performance found by the two methods. We present the median value (shown by line), and the interquartile range (the difference between the $3^{rd}$ and the $1^{st}$ quartile) shown by the shaded region. The three cases show that Augmented BO yields less search cost and reduces the variance. This demonstrates  Augmented BO is not fragile.

Another interesting observation is that
Augmented BO not only alleviates the fragile problem in \emph{Region II} but also
moves workloads from \emph{Region III} to \emph{Region II}.
~\myfigure{\ref{fig:convergence_time_1}} and
~\myfigure{\ref{fig:convergence_time_2}} are example workloads
in \emph{Region III}.
The first quartile indicates that Augmented BO finds the optimal configuration
even with four or five attempts in 25\% initial points that are tested.

\section{Practical Implications}\label{sec:practice}
\subsection{Bayesian Optimization in Practice}\label{sec:bo_practise}
In practice, users can tolerate a loss in performance (deployment cost or execution time) in exchange for lower search cost. In this section, we examine the performance of the two methods
when we (slightly) relax the definition of optimality.
Due to space limitations, we only present the results of minimizing deployment cost as we have shown it is more challenging, and the conclusion is similar to minimizing execution time.

To demonstrate the performance (of BO) and search cost trade-off, we vary the stopping criteria to understand how they affect both search cost and the best VM they find. 
We choose EI as the stopping criteria for Naive BO (as prescribed by \emph{CherryPick}). For Augmented BO, we use Prediction Delta and vary the thresholds from 0.9 to 1.3. We examine the three regions separately to analyze the effects of stopping criterion on different categories of workload.

In Figure~\ref{fig:stopping_criteria_comparison_good}, Naive BO finds the optimal VM regardless of the stopping criteria. This is counter-intuitive because there should exist a trade-off between the deployment cost and the search cost. We hypothesize that Naive BO cannot estimate that it has found the optimal VM. Augmented BO, on the other hand, clearly shows the trade-off. Augmented BO with the thresholds $1.25$ and $1.3$ performs similarly to Naive BO.
As pragmatic engineers, we are always hard-pressed to recommend Naive BO over Augmented BO, which achieves a performance of 1.04 rather than a perfect 1.0. 

In Figures~\ref{fig:stopping_criteria_comparison_bad} and~\ref{fig:stopping_criteria_comparison_problematic}, Augmented BO is the clear winner. With the $1.1$ threshold, Augmented BO outperforms Naive BO in both
the search cost and the deployment cost. To simplify the comparison, we choose 10\% EI for Naive BO (as prescribed by \emph{CherryPick}) and 1.1 threshold for Augmented BO. Our method yields lower search cost while achieving lower deployment cost. On average, it finds VMs which 5\% lower in deployment cost while reducing search cost by 20\%.
This demonstrates that the low-level augmented Bayesian Optimization finds the well-suited VMs quicker and is more precise when compared to Naive BO.

Overall, we recommend using $1.1$ threshold in Augmented BO since
the deployment cost is comparable with Naive BO and reduces
the search cost.
In \myfigure{\ref{fig:comparison_cost}},
we present the overall comparison of the two methods with the EI and threshold
described above. The horizontal axis represents the reduction in search cost, and the vertical axis represents the decrease in the deployment cost (higher the better in both).
The figure shows the result for all 107 workloads represented as points. Points enclosed with lines $x=0$ and $y=0$ (shown in \textcolor{blue}{blue} shade) indicates workloads, where Augmented BO can find VMs which have lower deployment cost using lower search cost. For example, the workload represented in (24, 10) is the case where Augmented BO uses 24\% lower search cost, and the best VM found (for that workload) has 10\% lower deployment cost than the one found by Naive BO. There are 46 such workloads.
Augmented BO requires higher search cost than Naive BO in five workloads (region shaded in \textcolor{red}{red}). But they both find the optimal solution.
There are 17 workloads where Augmented BO finds VM types with higher running cost
but with lower search cost---a region of trade-off.

\subsection{Time-Cost Trade-off}
\label{sec:tradeoff}

This section demonstrates how to adapt Augmented BO as well as Naive BO to navigate the time-cost trade-off. In practice, a user would always want a solution to reduce time as well as cost. We propose a new measure called time-cost product which is similar to an energy-time trade-off in high-performance computing~\cite{Freeh2007}.
Not every time-cost trade-off is desirable because
a small improvement in performance may incur a higher running cost.
For example, a 10\% improvement in execution time requires
a 50\% increase in deployment cost. 

For simplicity, we assign the same importance to time and cost.
That is, it is considered desirable for a 10\% improvement in time and
a 10\% increase in cost.
To support the time-cost trade-off, instead of predicting the execution time
and deployment cost, the surrogate model estimates the product of time and cost. Any two VMs are considered the same if their products of execution time and deployment cost are the same.
Similarly, a larger product represents an undesirable choice.

\begin{figure}
    \centering
    \includegraphics[width=0.4\textwidth]{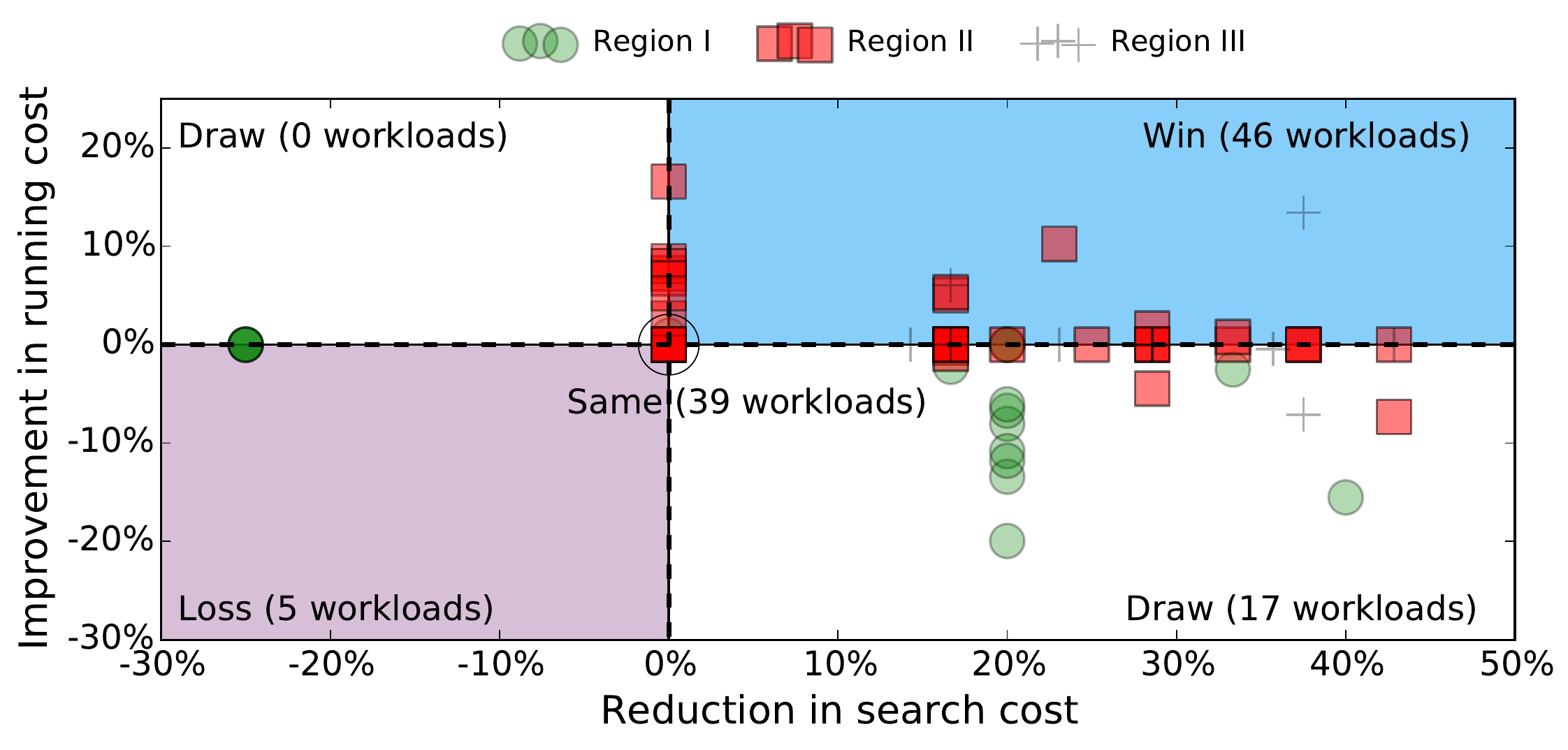}
    \vspace*{-4mm}
    \caption{Overall comparison for the two BO methods in finding the most cost-effective VM type across the evaluated 107 workloads. The numbers are calculated as the reduction percentage in search cost and improvement in deployment cost, both higher the better. Workloads in (0,0) represent workload which achieve similar performance in both methods.}
    \label{fig:comparison_cost}
    \vspace*{-6mm}
\end{figure}

\myfigure{\ref{fig:comparison_cdp_1.05}}  presents the comparison which is similar to Figure~\ref{fig:comparison_cost}. We observe a great reduction, \ie{$>50\%$} in search cost.
Naive BO exhibits long searching process (more than six attempts) in 24 percent workloads and very long searching (at least ten attempts) in 13 percent workloads. On the other hand, Augmented BO requires no more than 6 actual evaluation
for all 107 applications. Please note that the threshold used for this experiment is 1.05, which also tells us that the stopping criteria also need to be changed based on the workload as well as the performance objective.

%% file: relatedwork.tex
\section{Related Work}
\label{sec:related}

\noindent{\textbf{State-of-the-art}}:
\emph{Ernest} exploits the internal structure of a workload
to predict performance when running in different cluster sizes.
This greatly reduces the search cost because
a workload can be experimentally tested on a smaller cluster.
\emph{CherryPick} implements an optimization engine that uses
Bayesian Optimization in searching for
the best configuration~\cite{Alipourfard2017}.
However, \emph{CherryPick} does not leverage
low-level information and uses Gaussian Process-based BO, which makes it fragile.
\emph{PARIS} shares the same goal with our work~\cite{Yadwadkar2017}.
It builds a comprehensive performance model from the large training dataset and uses it along with current measurements
However, we argue that such approach might not be suitable for
batch processing workloads because of the low prediction accuracy
as discussed in Section~\ref{sec:smbo}.

\noindent \textbf{System Performance tuning}:
BOAT is a structured Bayesian Optimization-based framework for automatically tuning system performance
~\cite{Dalibard2017} which leverages contextual information.
BOAT combines the parametric and non-parametric model
for better predicting the trend in system performance.
The idea behind their work and our work is very similar:
leveraging domain knowledge to enhance BO.

Bilal et al. propose a framework to automate tuning system performance
of stream-processing systems.Their modified hill-climbing search with heuristic sampling
inspired by Latin Hypercube improves the search process by two to five times.
Several papers use minimal sampling techniques to build models to optimize software systems \cite{nair2017using},~\cite{oh2017finding, nair2017flash, nair2017faster}. 
The above methods reduce the search cost by a significant degree.
However, they focus on performance tuning for the same workload (or application)
on the same type of machine. It is not clear how to leverage their approaches to support different
machine configurations in cloud computing. We, instead, find the best machine configuration for a given workload.


\noindent \textbf{Leveraging low-level performance}:
Low-level performance information is leveraged
to identify performance bottlenecks and to predict application performance.
DeepDive is designed to identify performance interference of co-existing VMs
~\cite{Novakovic2013}.
Wang et al. propose using the CART model to predict storage performance.
Their approach requires workload information, which may not be practical
for our problem setting.
Inside-out provides reliable performance prediction of distributed storage
service by using only low-level performance information~\cite{Hsu2016}.
The authors show that high-level performance can be accurately captured
by only the low-level metrics.
This accurate prediction model can be used to adjust resource allocation
for meeting performance objectives.

\begin{figure}
    \centering
    \includegraphics[width=0.4\textwidth]{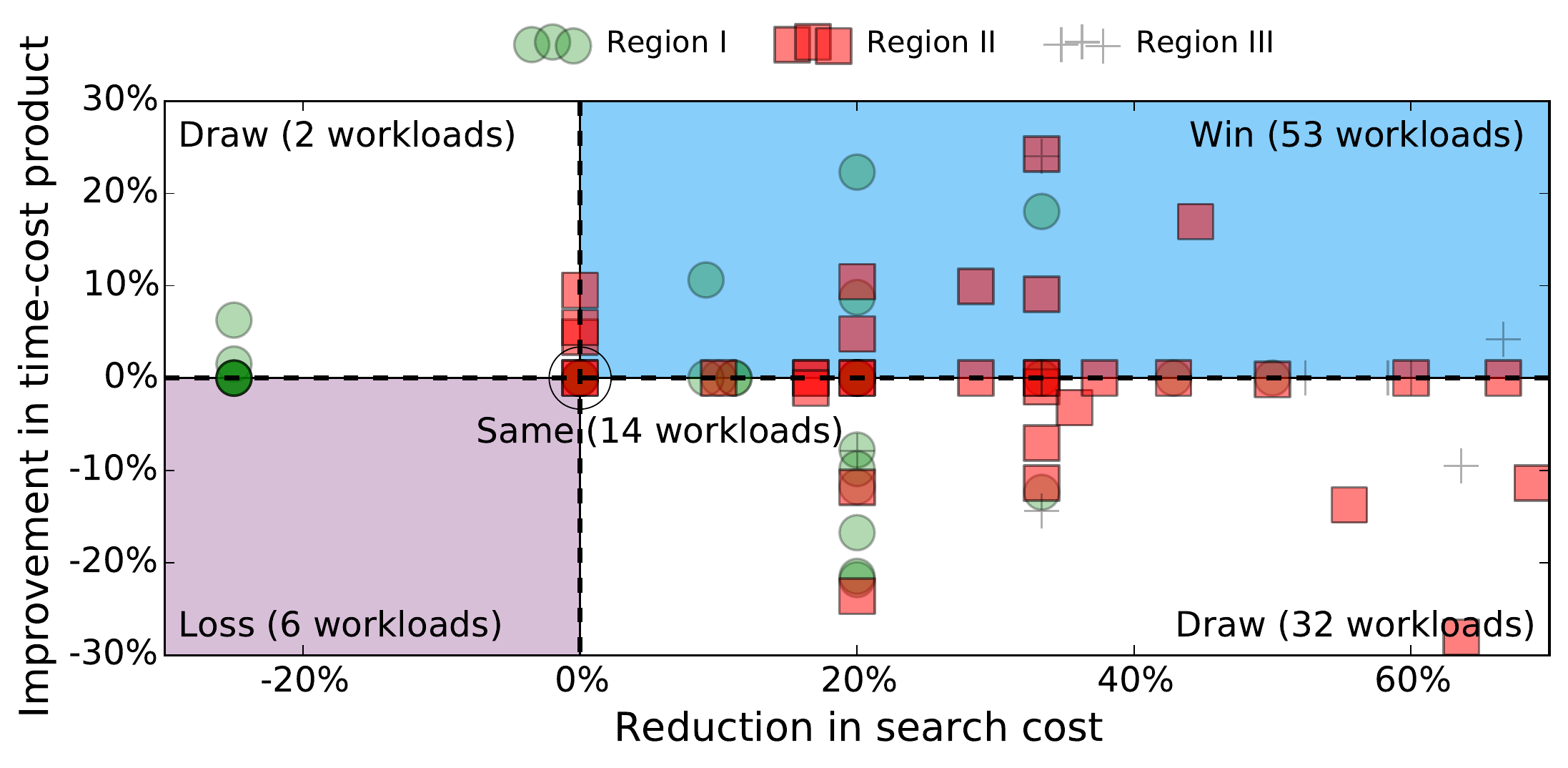}
    \vspace*{-4mm}
    \caption{Similar to \myfigure{\ref{fig:comparison_cost}}, the optimization objective is to find the best configuration both in execution time and search cost. Augmented BO supports finding the best VM type, given a time-cost tradeoff.}
    \label{fig:comparison_cdp_1.05}
    \vspace*{-6mm}
\end{figure}

%% file: conclusion.tex
\section{Conclusion}
\label{sec:conclusion}

In this paper, we identify and demonstrate the fragility of Bayesian Optimization in finding the best cloud VM type. The fragility arises from the inadequate information used to represent the instance space. This fragility affects prior work which uses only instance space to guide Bayesian Optimization. To overcome the problem of fragility, we augment the instance space with low-level performance information, which is known to be useful for characterizing the performance of the system. We present our method, Augmented Bayesian Optimization, which seamlessly integrates the low-level metrics (obtained with negligible overhead) to the surrogate model. Additionally, we make design choices to modify existing BO to make more informed decisions. We demonstrate empirically the usefulness of Augmented BO by showing that Augmented BO can find the best VM type across all workloads. In 46 out of 107 workloads, Augmented BO outperforms the state-of-the-art Bayesian optimization method in terms of both performance and search-cost.  

More generally, we conclude that it is often insufficient to use general-purpose off-the-shelf methods (BO in this case) for selecting the best VM without augmenting those methods with essential systems knowledge such as CPU utilization, working memory size and I/O wait time.  In our future work, we plan to further augment Bayesian Optimizer with historical performance data to further reduce the search cost.